\documentclass[titlepage,aps,prf,superscriptaddress]{revtex4-2}
\usepackage{graphicx}
\usepackage{picture}
\usepackage{amsmath}
\usepackage{hyperref}
\usepackage{bigints}
\usepackage[]{color}

\newcommand{\de}{\mathrm{d}}

\begin{document}


\title{Influence of density and viscosity on deformation,\\ breakage and coalescence of bubbles in turbulence}


\author{Francesca Mangani}
\affiliation{Institute of Fluid Mechanics and Heat Transfer, TU-Wien, 1060 Vienna, Austria}
\author{Giovanni Soligo}
\affiliation{Institute of Fluid Mechanics and Heat Transfer, TU-Wien, 1060 Vienna, Austria}
\affiliation{Complex Fluids and Flows Unit, OIST, 904-0495 Okinawa, Japan}
\altaffiliation{Now at Complex Fluids and Flows Unit, OIST, 904-0495 Okinawa, Japan}
\author{Alessio Roccon}
\affiliation{Institute of Fluid Mechanics and Heat Transfer, TU-Wien, 1060 Vienna, Austria}
\affiliation{Polytechnic Department, University of Udine, 33100 Udine, Italy}
\author{Alfredo Soldati}
\affiliation{Institute of Fluid Mechanics and Heat Transfer, TU-Wien, 1060 Vienna, Austria}
\affiliation{Polytechnic Department, University of Udine, 33100 Udine, Italy}


\date{\today}

\begin{abstract}
We numerically investigate the effect of density and viscosity differences on a swarm of large and deformable bubbles dispersed in a turbulent channel flow. 
For a given shear Reynolds number, $Re_\tau=300$, and a constant bubble volume fraction, $\Phi \simeq5.4\%$, we perform a campaign of direct numerical simulations (DNS) of turbulence coupled with a phase-field method (PFM) accounting for interfacial phenomena.
For each simulation, we vary the Weber number ($We$, ratio of inertial to surface tension forces), the density ratio ($\rho_r$, ratio of bubble density to carrier flow density) and the viscosity ratio ($\eta_r$, ratio of bubble viscosity to carrier flow viscosity).
Specifically, we consider two Weber numbers, $We=1.50$ and $We=3.00$, four density ratios, from $\rho_r=1$ down to $\rho_r=0.001$ and five viscosity ratios from $\eta_r=0.01$ up to $\eta_r=100$.
Our results show that density differences have a negligible effect on breakage and coalescence phenomena, while a much stronger effect is observed when changing the viscosity of the two phases.
Increasing the bubble viscosity with respect to the carrier fluid viscosity damps turbulence fluctuations, makes the bubble more rigid and strongly prevents large deformations, thus reducing the number of breakage events.
Local deformations of the interface, on the contrary, depend on both density and viscosity ratios: as the bubble density is increased, a larger number of small-scale deformations, small dimples and bumps, appear on the interface of the bubble.
The opposite effect is observed for increasing bubble viscosities: the interface of the bubbles become smoother.
We report that these effects are mostly visible for larger Weber numbers, where surface forces are weaker.
Finally, we characterize the flow inside the bubbles; as the bubble density  is increased, we observe, as expected, an increase in the turbulent kinetic energy (TKE) inside the bubble, while as the bubble viscosity is increased, we observe a mild reduction of the TKE inside the bubble and a strong suppression of turbulence.



\end{abstract}



\maketitle

\graphicspath{{./figures/}}

\section{Introduction}

Interactions among turbulence and deformable interfaces are common in many physical instances, from ocean waves formation \cite{melville1996role,jessup1997infrared} to atomization processes \cite{gorokhovski2008modeling}, as well as drops and bubbles entrained in a turbulent flow \cite{mathai2020bubbly,Elghobashi2018,soligo2021}.
The outcome of these interactions is of fundamental importance as it controls the exchanges of heat, mass, and momentum across the interface and thus between the two phases.
The study of  turbulence-interface interactions, however, is a non-trivial task as these interactions are governed by a physics acting at very different spatio-temporal scales:  from the largest problem scale, down to the Kolmogorov scale of turbulence and further down to the molecular scale of the interface.
This multi-scale nature makes the investigation of multiphase turbulence very challenging.
In particular, experimental investigations using optical techniques are usually limited to small volume fractions due to the difficulty of accessing phases with heterogeneous optical properties \cite{boyer2002measuring,deen2002two,poelma2006turbulence} and limited range of length scales that can be possibly measured.
In this scenario, despite some limitations, numerical simulations represent an essential tool to investigate multiphase flows as they allow to access detailed space- and time-resolved information on the flow field and dispersed phase.
Specifically, direct numerical simulation (DNS), in which all relevant scales of turbulence are resolved, proved to be a tool of paramount importance for a deeper understanding of single-phase \cite{rogallo1984numerical,Kim1987} and multiphase turbulence \cite{mathai2020bubbly,Elghobashi2018,soligo2021}.

In this work we focus on the interactions of a swarm of large and deformable bubbles or drops (bubbles hereinafter without any loss of generality) with wall-bounded turbulence (turbulent channel flow).
This setup has been widely used in the past to investigate different aspects of bubbly flows, from bubbles shape, deformation  and clustering to the flow modifications produced by the bubbles themselves.
In the pioneering works of Lu \& Tryggvason \cite{Lu2007,Lu2008}, the effects of the bubble size and deformability were investigated: they observed that as bubbles become more deformable, they move towards the middle of the channel and have a relatively small effect on the flow-rate. 
Scarbolo {\em et al.} \cite{Scarbolo2015,Scarbolo2016}, considering a matched density and viscosity system, investigated the effect of the surface tension, observing that surface tension forces play a key role in determining the dispersed phase topology.
Roccon {\em et al.} \cite{Roccon2017} studied the effect of the bubble viscosity, finding that for small surface tension values, larger internal viscosities reduce the drop deformability. 
Recently, Soligo {\em et al.} \cite{Soligo2019c,soligo2020effect}, considering  also the presence of a soluble surfactant, investigated the surfactant effects on drop morphology  \cite{Soligo2019c} and flow behavior \cite{soligo2020effect}.
Finally, Hasslberger {\em et al.} \cite{hasslberger2020direct} analyzed the coherent structures obtained in a bubble-laden turbulent channel flow while Cannon {\em et al.} \cite{cannon2021effect} investigated the role played by droplets coalescence on drag in turbulent channel flows.

The foremost goal of this paper is  to improve the fundamental understanding of bubble-bubble and bubble-turbulence interactions.
Indeed, bubbles transported by a turbulent flow are characterized by complex dynamics, as they collide, coalesce and break apart.
This behavior is governed by the forces generated by the surrounding continuous phase, acting on the surface of the bubbles with shear and normal stresses, and by the response of bubbles, which depends on their surface tension and their density and viscosity. 
The ultimate competition among these forces determines the number, shape, and deformation of the bubbles.
In this work, we want to extend our previous works \cite{Scarbolo2015,Roccon2017} and provide a comprehensive analysis on the effects of density ratio (ratio between the density of the bubble phase over the dispersed phase), viscosity ratio (ratio between the viscosity of the bubble phase over the dispersed phase) and surface tension (controlled by the Weber number, ratio of inertial over surface tension contributions) on the multiphase system.
Specifically, the first objective is to investigate the effects of these parameters on the dispersed phase topology and its topological modifications (coalescence and breakage events), and to characterize the shape and deformation of the bubbles.
The second objective of this work is to characterize the global and local flow modifications produced by bubbles on the turbulent channel flow behavior.
To this aim, we build and analyze a database of direct numerical simulations of turbulent channel flows laden with deformable bubbles, considering different values of  density ratios, viscosity ratios, and surface tension.
The numerical framework of the simulations relies on a direct solution of the Navier-Stokes equations coupled with a phase-field method.
Direct solutions of the Navier-Stokes equations are used to accurately resolve all the relevant turbulence scales, while the phase-field method \cite{Jacqmin1999,Badalassi2003} -- an interface capturing approach that relies on an order parameter to define the local concentration of each phase -- is used to describe in a thermodynamically consistent manner the motion of the deformable interface and its topological modifications (i.e. coalescence and breakage events).

The paper is organized as follows: in section~\ref{sec: method}, we introduce the numerical method, the simulation setup, and the parameters of the simulations.
Then, in section~\ref{sec: results}, we present the results obtained from the analysis of the simulations database.
First, we focus on the effects of density and viscosity ratios and surface tension values on the topology of the dispersed phase and its topological changes (breakage and coalescence).
Secondly, we evaluate the effects of these parameters on the overall interfacial area and curvature of the bubble interface. 
Thirdly, we study the effects of density and viscosity ratios and Weber number on the mean velocity profiles and on the turbulent kinetic energy (TKE) of the bubbles.
Finally, we summarize the results and draw our conclusions in section~\ref{sec: conclusions}.

\section{Methodology}
\label{sec: method}

We consider the case of a swarm of bubbles injected in a turbulent channel flow with a rectangular cross-section.
The dispersed and carrier phases are characterized by density $\rho_d$  and $\rho_c$, and viscosity $\eta_d$ and $ \eta_c$, where the subscripts $d$ and $c$ identify the dispersed and carrier phase, respectively.
We define the density ratio and viscosity ratio as $\rho_r=\rho_d/\rho_c$ and $\eta_r=\eta_d/\eta_c$ respectively.
The interface that separates the two phases is characterized by a constant and uniform value of the surface tension, $\sigma$. 
To describe the dynamics of the system, direct numerical simulation (DNS) of the Navier-Stokes equations, used to describe the flow field, are coupled with a phase-field method (PFM), used to describe interfacial phenomena \citep{Jacqmin1999,Badalassi2003}.

\subsection{Modeling of interfacial phenomena}

The phase-field method uses an order parameter, the phase field $\phi$, to identify the two phases: the order parameter is uniform in the bulk of each phase ($\phi= \pm 1$) and undergoes a smooth transition across the interface. Indeed, the sharp interface is replaced by a thin transition layer.
The transport of the phase field $\phi$ is described by the Cahn-Hilliard equation, which in dimensionless form reads as:
\begin{equation}
\frac{\partial \phi}{\partial t}+\mathbf{u} \cdot \nabla \phi=\frac{1}{Pe}\nabla^2 \mu + f_p \, ,
\label{eq:ch}
\end{equation}
where ${\bf u}=(u,v,w)$ is the velocity vector, $Pe$ is the P\'eclet number, $\mu$ is the chemical potential and $f_p$ is the penalty flux introduced with the profile-corrected formulation of the phase-field method \cite{li2016phase,zhang2017,Soligo2019b}.
The P\'eclet number is defined as follows:
\begin{equation}
Pe=\frac{u_\tau h}{\mathcal{M} \beta} \, ,
\end{equation}
where $u_\tau=\sqrt{\tau_w/\rho_c}$ is the friction velocity (being $\tau_w$ the shear stress at the wall and $\rho_c$ the carrier phase density), $h$ is the channel half-height, $\mathcal{M}$ is the mobility parameter and $\beta$ is a positive constant introduced to make the chemical potential dimensionless.
The P\'eclet number identifies the ratio between the diffusive time-scale, $h^2/\mathcal{M} \beta$, and the convective time-scale, $h/u_\tau$, of the interface.

The chemical potential $\mu$ is defined as the variational derivative of a Ginzburg-Landau free-energy functional, the expression of which is selected to represent an immiscible binary mixture of isothermal fluids \citep{Soligo2019a,Soligo2019b,Soligo2019c}.
The functional is composed by the sum of two different contributions: the first contribution, $f_0$, accounts for the tendency of the system to separate into the two pure stable phases, while the second contribution, $f_{mix}$, is a mixing term accounting for the energy stored at the interface.    
The mathematical expression of the functional is:
\begin{equation}
\mathcal{F}[\phi, \nabla \phi]=\bigintsss_{\Omega} \bigg( \underbrace{\frac{(\phi^2-1)^2}{4}}_{f_0}+\underbrace{\frac{Ch^2}{2} \left| \nabla \phi \right|^2}_{f_{mix}} \bigg)  \de \Omega \, ,
\label{ginz-land}
\end{equation}
where $\Omega$ is the domain considered and $Ch$ is the Cahn number, which represents  the dimensionless thickness of the thin interfacial layer between the two fluids ($\xi$ is the physical thickness of the interface).
\begin{equation}
Ch=\frac{\xi}{h} 
\end{equation}
From equation~(\ref{ginz-land}), the expression of the chemical potential can be derived as the functional derivative with respect to the order parameter:
\begin{equation}
\mu=\frac{\delta \mathcal{F}[\phi, \nabla \phi ]}{\delta \phi}=\phi^3 - \phi - Ch^2  \nabla^2 \phi \, .
\end{equation}
At the equilibrium, the chemical potential will be constant throughout all the domain. 
The equilibrium profile for a flat interface can thus be obtained solving $\nabla \mu =0$, hence obtaining:
\begin{equation}
\phi_{eq}=\tanh \left( \frac{s}{\sqrt{2}Ch}\right)
\end{equation}
where $s$ is a coordinate normal to the interface.

Finally, $f_p$ is the penalty-flux employed in the profile-corrected formulation of the phase-field method.
This formulation is an improvement to the standard phase-field formulation: it allows to better maintain the equilibrium interfacial profile and it overcomes the drawbacks of the method (e.g. mass leakages among the phases and misrepresentation of the interfacial profile \cite{YUE2007,li2016phase}).
This penalty flux is defined as:
\begin{equation}
f_p=\frac{\lambda}{Pe} \left[ \nabla^2 \phi -\frac{1}{\sqrt{2}Ch} \nabla \cdot  \left( (1-\phi^2) \frac{\nabla\phi}{|\nabla\phi|} \right) \right] \, ,
\end{equation}
where the numerical parameter $\lambda$ can be set via the scaling $\lambda=0.0625/Ch$ \cite{Soligo2019b}. 

Before proceeding, it is worth to briefly discuss the main capabilities and limitations of interface-resolved simulations in describing topological modifications of the interface \cite{tryggvason2013multiscale,Soligo2019c,soligo2021}.
The numerical description of breakages and coalescences is indeed one of the most challenging aspects of interface-resolved simulation methods.
A fully-resolved simulation of topological changes would require resolving all the scales, from the molecular scale of the interface \cite{rekvig2007} up to the largest scales of the flow. 
This type of simulation, however, is way beyond the capabilities of any existing supercomputing facility. 
The common choice is to avoid resolving the small interfacial scales and to find a way to approximate their dynamics on a much larger scale. 
Here, following a similar approach, we limit the resolved range to the scales of turbulence: from the Kolmogorov length scale up to the problem size. 
Thus, phenomena occurring at scales smaller than Kolmogorov are smeared out on the smallest resolved scale. 
This choice however influences the description of coalescence and breakage events.
For coalescences, a part of the physics involved in the coalescence process \cite{kamp2017drop} (i.e. film drainage and rupture) cannot be directly resolved. 
As a result, regardless of the approach employed to describe coalescence (models for interface tracking methods \cite{Tryggvason2001,Lu2018} or implicit description for interface capturing methods  \cite{Dodd2016,soligo2021}), numerical simulations struggle in predicting physical coalescence, with this inaccuracy referred to as numerical coalescence. 
For breakages, the picture is different and their numerical description is less troublesome.
Indeed, being breakage a very quick phenomenon, it can be well approximated without resolving the dynamics at the molecular scale and there is evidence that the Navier-Stokes equations alone provide an adequate description of a breakage event \cite{eggers1995}. 
Besides, the small time scale of a breakage limits the impact of the approximation on the overall flow dynamics \cite{Herrmann2011,Lu2018}.
Therefore, the description of breakages on turbulence-resolved grids is considered to be rather accurate, although in the pinch-off region the smallest interfacial features, characterized by high curvature, may not be perfectly resolved.

\subsection{Hydrodynamics}
\label{hydro}

To describe the hydrodynamics of the multiphase system, the Cahn-Hilliard equation is coupled with the Navier-Stokes equations.
The presence of a deformable interface (and of the corresponding surface tension forces) is accounted for by introducing an interfacial term in the Navier-Stokes equations. 
Recalling that in the present study we consider two fluids having different densities and viscosities, we use here the formulation of continuity and Navier-Stokes equations proposed by Dong \& Shen \cite{dong2012time}.
The resulting governing equations for the hydrodynamics read as follow:
\begin{equation}
\nabla \cdot \mathbf{u}=0 \, ,
\label{eq:cont}
\end{equation}
\begin{equation}
\rho (\phi) \left( \frac{\partial \mathbf{u}}{\partial t}+\mathbf{u} \cdot \nabla \mathbf{u} \right)=
-\nabla p +\frac{1}{Re_\tau} \nabla \cdot [ \eta (\phi) ( \nabla \mathbf{u} + \nabla \mathbf{u}^{T})]+ \frac{3}{\sqrt{8}}\frac{Ch}{We} \nabla \cdot \mathsf{T_c}  \, ,
\label{eq:ns}
\end{equation}
where ${\bf u}=(u,v,w)$ is the velocity vector, $p$ is the pressure field, $\mathsf{T_c}$ is the Korteweg tensor and ${\rho}(\phi)$ and ${\eta}(\phi)$ are the density and viscosity fields, respectively.
The density and viscosity fields are dimensionless scalar functions that account for the local value of density and viscosity respectively \citep{alpak2016phase,DING2007,Kim2012}; the carrier phase properties are used to make these fields dimensionless.
The local density and viscosity are assumed to be linear functions of the phase field:
\begin{equation}
\rho(\phi)=  1 + (\rho_r -1) \frac{\phi + 1}{2} \, ,
\end{equation}
\begin{equation}
\eta(\phi)=  1 + (\eta_r -1) \frac{\phi + 1}{2}  \, ,
\end{equation}
where $\rho_r$ and $\eta_r$ are the density and viscosity ratios, respectively.

The Korteweg tensor \citep{KORTEWEG1901}, used to account for the surface tension forces, is defined as follows:
\begin{equation}
\mathsf{T_c}=|\nabla\phi|^2 \mathsf{I}-\nabla\phi\otimes \nabla \phi \, .
\end{equation}

The dimensionless groups appearing in the Navier-Stokes equations are the shear Reynolds number, $Re_\tau$, and the Weber number, $We$, which are defined as:
\begin{equation}
Re_\tau=\frac{\rho_c u_\tau h}{\eta_c}\, ,  \qquad
We=\frac{\rho_c u_\tau^2 h}{\sigma}\,.
\label{dp}
\end{equation}
The Reynolds number represents the ratio between inertial and viscous forces, while the Weber number is the ratio between inertial and surface tension forces.
Both Reynolds and Weber numbers are defined using the carrier phase properties ($\rho_c$ and $\eta_c$).

\subsection{Numerical method}

The governing equations~(\ref{eq:ch}), (\ref{eq:cont}) and~(\ref{eq:ns}) are solved using a pseudo-spectral method, which uses Fourier series along the periodic directions (streamwise and spanwise) and Chebyshev polynomials along the wall-normal direction.
The Navier-Stokes and continuity equations are solved using the velocity-vorticity formulation: equation~(\ref{eq:ns}) is rewritten as a $4^{th}$ order equation for the wall-normal component of the velocity $u_z$ and a $2^{nd}$ order equation for the wall-normal component of the vorticity $\omega_z$ \citep{Kim1987,Speziale1987}. 
Equation~(\ref{eq:ch}) is also split into two $2^{nd}$ order equations \citep{Badalassi2003}; this way the governing equations are recasted as a coupled system of Helmholtz equations, which can be readily solved.
The governing equations are time advanced using an implicit-explicit scheme. 
For the Navier-Stokes equations, the non-linear term is first rewritten as the sum of a linear and a non-linear contribution \citep{Zonta2012}.
Then, the linear part is integrated using a Crank-Nicolson implicit scheme, while the non-linear part is integrated using an Adams-Bashforth explicit scheme.
Likewise, for the Cahn-Hilliard equation, the linear term is integrated using an implicit Euler scheme, while the non-linear term is integrated in time using an Adams-Bashforth scheme. 
The adoption of the implicit Euler scheme helps damping unphysical high-frequency oscillations that could arise from the steep gradients of $\phi$.

\subsection{Boundary conditions}

The resulting set of governing equations is  complemented by suitable boundary conditions.
For the Navier-Stokes equations, no-slip boundary conditions are enforced at the top and bottom wall ($z/h=\pm 1$):
\begin{equation}
{\bf u} (z/h=\pm 1)=0\, .
\end{equation}
For the Cahn-Hilliard equation,  no-flux boundary conditions are applied at the two walls, yelding the following boundary conditions:
\begin{equation}
\frac{\partial \phi}{\partial z}(z/h=\pm 1)=0 \, , \qquad
\frac{\partial^3 \phi}{\partial z^3}(z/h=\pm 1)=0 \, .
\end{equation}
Along the streamwise and spanwise directions ($x$ and $y$), periodic boundary conditions are imposed for all variables (Fourier discretization).
The adoption of these boundary conditions leads to the conservation of the phase field over time:
\begin{equation}
\frac{\partial}{\partial t} \int_{\Omega}^{} \phi \de \Omega = 0\, .
\end{equation}
This enforces mass conservation of the entire system but does not guarantee the conservation of the mass of each phase \citep{YUE2007,Soligo2019b}, as some leakages between the phases may occur. 
This drawback is rooted in the phase-field method and is here mitigated with the adoption of the profile-corrected formulation.
In the present cases, mass leakages are limited to at most 8\% of the dispersed phase mass and occur only in the initial transient phase; once the statistically-stationary condition is reached, the mass of each phase keeps constant. 

\subsection{Simulation set-up}
\label{sec: setup}

We consider a turbulent channel flow at a shear Reynolds number $Re_\tau=300$ for all the cases.
The computational domain has dimensions $L_x \times L_y \times L_z= 4\pi h   \times 2\pi h \times 2h$, which corresponds to $L_x^+ \times L_y^+ \times L_z^+=3770 \times 1885 \times 600$ wall units.
The domain is discretized with $N_x \times N_y \times N_z=512\times 256 \times513$ grid points; the computational grid has uniform spacing in the homogenous directions, while Chebyshev-Gauss-Lobatto points are used in the wall-normal direction.
The flow is driven by an imposed constant pressure gradient in the streamwise direction. 
We consider two surface tension values, which are set via the Weber number: $We=1.50$ (higher surface tension) and $We=3.00$ (lower surface tension). 
The selected values are characteristics of air/water mixtures \cite{jasper1972surface}.
For each surface tension value (i.e. for each Weber number), we first keep a unitary density ratio and we analyze the effect of different viscosity ratios: from $\eta_r=0.01$ (less viscous bubbles) up to $\eta_r=100$ (more viscous bubbles).
Then, we keep a unitary viscosity ratio and we consider different density ratios: from $\rho_r=1$ (matched density bubbles) down to $\rho_r=0.001$ (lighter bubbles).
Finally, to evaluate the combined effect of density and viscosity differences, we consider a case in which both bubble density and viscosity are smaller than those of the carrier fluid: $\rho_r=0.1$ and $\eta_r=0.1$.
In addition, we perform a single-phase flow simulation as a reference case and to provide initial velocity fields for the multiphase simulations.
It is worthwhile noting that when different properties (i.e. density and viscosity) are considered, the local value of the Reynolds number changes as well as the range of spatiotemporal scales that needs to be resolved to fulfill the DNS requirements.
These modifications can be appreciated from table~\ref{tab:grid} in which we show an estimate of the turbulence length scale inside the dispersed phase (computed from the definition of the Kolmogorov length scale), $\eta_{k,d}^+$, the grid resolution, the final average bubble-size, $\langle d_{eq}^+ \rangle$, and its root mean square value, RMS($d_{eq}^+)$, for all the different combination of density and viscosity ratios considered as well as for the reference single-phase case.
The bubble size has been characterized using the equivalent diameter, $d_{eq}^+$,  i.e. the diameter of an equivalent spherical bubble with the same volume as the bubble considered \cite{Soligo2019c}:
\begin{equation}
d_{eq}^+=\left( \frac{6 V^+}{\pi} \right)^{1/3}
\end{equation}
where $V^+$ is the volume of the bubble.
All dimensions are reported in wall units (based on the carrier flow shear Reynolds number) and refer to the channel centre, where most bubbles are located.
The Kolmogorov scale, which is used here to provide an estimate of the smallest length scale inside the bubbles, has been computed as follows:
\begin{equation}
\eta_{k,d}^+=\left( \frac{\eta_r^2 Re_\tau^2}{\epsilon^+} \right)^{1/4}
\end{equation}
where $\epsilon^+$ is the dissipation at the channel center evaluated in the region characterized by $\phi \ge 0$ (i.e. inside the bubbles), $\eta_r$ is the viscosity ratio and $Re_\tau$ is the shear Reynolds number.
We can observe that for almost all the cases presented here, the estimated Kolmogorov scale is of the order of the grid spacing thus ensuring a correct resolution of all the relevant flow scales.
Only for the cases with $\eta_r \leq 0.1$ (most critical cases due to the largest local Reynolds number increase), the smallest flow scales (which are found inside the bubbles) cannot be completely resolved.
From table~\ref{tab:grid}, we can also observe that the average bubble size is always at least one order of magnitude larger than the grid spacing.

\begin{ruledtabular}
\begin{table}[!h]
\centering
\begin{tabular}{cccccccccc}
\hline 
System&$We$ & $\eta_r$ & $\rho_r$ & $\Delta x^+$& $\Delta y^+$ & $\Delta z^{+}$ & $\eta_{k,d}^+$ & $\langle d_{eq}^+ \rangle $& RMS($d_{eq}^+$)\\
\hline 
\hline
Single-phase& -    &   -   &      -    & 7.36 & 7.36 & 1.84&   4.19    &   - & -\\
\hline
Bubbles-laden &1.50  &0.01 & 1.000 & 7.36 & 7.36 & 1.84 & 0.20    &195.13& 176.97\\
Bubbles-laden &1.50  &0.10 & 1.000 & 7.36 & 7.36 & 1.84 & 1.04    &191.16& 134.04\\
Bubbles-laden &1.50  &1.00 & 1.000 & 7.36 & 7.36 & 1.84 &  5.27   &226.72& 123.55\\
Bubbles-laden &1.50  &10.0 & 1.000 & 7.36 & 7.36 & 1.84 & 26.72   &229.84& 127.99\\
Bubbles-laden &1.50  &100. & 1.000 & 7.36 & 7.36 & 1.84 &145.50 &245.04& 104.51\\
Bubbles-laden &1.50  &1.00 & 0.001 & 7.36 & 7.36 & 1.84 & 887.50 &208.15& 150.61\\
Bubbles-laden &1.50  &1.00 & 0.010 & 7.36 & 7.36 & 1.84 & 185.80 &230.31& 142.16\\
Bubbles-laden &1.50  &1.00 & 0.100 & 7.36 & 7.36 & 1.84 & 30.66 &180.60& 142.73\\
Bubbles-laden &1.50  &0.10 & 0.100 & 7.36 & 7.36 & 1.84 & 5.86  &186.00& 138.08\\
\hline
Bubbles-laden &3.00  & 0.01& 1.000 & 7.36 & 7.36 & 1.84 & 0.19  & 81.37 & 74.77\\
Bubbles-laden &3.00  & 0.10& 1.000 & 7.36 & 7.36 & 1.84 & 0.94  & 84.06 & 76.15\\
Bubbles-laden &3.00  & 1.00& 1.000 & 7.36 & 7.36 & 1.84 & 4.87  & 87.56 & 79.55\\
Bubbles-laden &3.00  & 10.0& 1.000 & 7.36 & 7.36 & 1.84 & 24.96 &89.70 & 77.94\\
Bubbles-laden &3.00  & 100.& 1.000 & 7.36 & 7.36 & 1.84 & 140.3 &203.62&111.09\\
Bubbles-laden &3.00  & 1.00 & 0.001& 7.36 & 7.36 & 1.84 & 818.2 &87.58 & 77.74\\
Bubbles-laden &3.00  & 1.00 & 0.010& 7.36 & 7.36 & 1.84 & 142.0 &86.54 & 76.25\\
Bubbles-laden &3.00  & 1.00 & 0.100& 7.36 & 7.36 & 1.84 & 27.45 & 91.16 & 81.28\\
Bubbles-laden &3.00  & 0.10 & 0.100& 7.36 & 7.36 & 1.84 & 4.63   &83.62 & 75.41\\
\hline
\end{tabular}
\caption{Grid resolution, $\Delta x^+$, $\Delta y^+$ and $\Delta z^+_c$, Kolmogorov scale at the channel centre in the dispersed phase, $\eta_{k,d}^+$, average equivalent diameter of the bubbles, $\langle d_{eq}^+ \rangle$, and root mean square of the bubble equivalent diameter, RMS($d_{eq}^+$), for all the different simulations performed.
All dimensions are reported in wall units; Kolmogorov scale is measured at the channel centre. 
Single-phase flow values at the channel centre have been also reported as a reference.}
\end{table}
\label{tab:grid}
\end{ruledtabular}


For the phase field, the Cahn number is set to $Ch=0.02$. 
This value is selected based on the grid resolution: at least three grid points are required across the interface to accurately describe the steep gradients present \cite{Soligo2019a}. 
The phase field P\'eclet number has been set according to the scaling $Pe=1/Ch=50$, to achieve convergence to the sharp interface limit \cite{YUE2007,Magaletti2013}.
More refined grids allow to reduce the thickness of the interface and to adopt smaller Cahn numbers.
However, the resulting computational cost is much larger: grid resolution needs to be refined along all three directions, as the orientation of the interfacial layer is arbitrary, and the time step has to be reduced as well to satisfy the Courant–Friedrichs–Lewy condition.
Overall, the computational cost of a simulation with an halved Cahn number is roughly 16 times larger: grid refinement makes the simulation eight times more expensive and the time step limitation makes the simulation twice as expensive.

At the beginning of each simulation, a regular array of 256 spherical droplets with diameter $d=0.4h$ (corresponding to $d^+=120$ wall units) is initialized in a fully-developed single-phase turbulent channel flow. 
The total volume fraction of the dispersed phase is $\Phi=V_d/(V_c+V_d)\simeq5.4\%$, being $V_d$ and $V_c$ the volume of the dispersed and carrier phase, respectively. 
As the array of spherical bubbles is suddenly released in a single-phase turbulent flow, turbulent velocity fluctuations strongly perturb the interfacial profile; during this initial coupling phase, mass leakages among the phases may occur \cite{YUE2007,Soligo2019b}
After this initial transient, the mass of each phase keeps constant over time.
While the initial condition chosen for the dispersed phase may seem unphysical, after a short transient, memory of the initial condition is completely lost and the results are not affected by the initial condition selected \cite{Soligo2019c}·
Different initial conditions have been tested (e.g., the injection of a thin liquid sheet at the channel center) and the same statistically statistically-stationary results were obtained.
We selected the current initial configuration as it reduces the time required to reach statistically-stationary conditions.
\begin{ruledtabular}
\begin{table}[!h]
\centering
\begin{tabular}{c cccccc}
\hline
System& $Re_\tau$  & $We$ & $\eta_r$ & $\rho_r$&$Ch$&$Pe$ \\
\hline
Single-phase    &300& -&-&-&-&- \\
\hline
Bubbles-laden &300&1.50&0.01&1.000&0.02&50 \\
Bubbles-laden &300&1.50&0.10&1.000&0.02&50\\
Bubbles-laden &300&1.50&1.00&1.000&0.02&50\\
Bubbles-laden &300&1.50&10.0&1.000&0.02&50\\
Bubbles-laden &300&1.50&100.&1.000&0.02&50\\
Bubbles-laden &300&1.50&1.00&0.001&0.02&50\\
Bubbles-laden &300&1.50&1.00&0.010&0.02&50\\
Bubbles-laden &300&1.50&1.00&0.100&0.02&50\\
Bubbles-laden &300&1.50&0.10&0.100&0.02&50\\
\hline
Bubbles-laden &300&3.00&0.01&1.000&0.02&50\\
Bubbles-laden &300&3.00&0.10&1.000&0.02&50\\
Bubbles-laden &300&3.00&1.00&1.000&0.02&50\\
Bubbles-laden &300&3.00&10.0&1.000&0.02&50\\
Bubbles-laden &300&3.00&100.&1.000&0.02&50\\
Bubbles-laden &300&3.00&1.00&0.001&0.02&50\\
Bubbles-laden &300&3.00&1.00&0.010&0.02&50\\
Bubbles-laden &300&3.00&1.00&0.100&0.02&50\\
Bubbles-laden &300&3.00&0.10&0.100&0.02&50\\
\end{tabular}
\caption{Overview of simulations parameters. 
Wa analyze two Weber numbers: $We=1.50$ and $We=3.00$.
For each Weber number, we consider four density ratios: from $\rho_r=0.001$ up to $\rho_r=1.000$; five viscosity ratios: from $\eta_r=0.01$ up to $\eta_r=100$ and a combined case $\rho_r=0.1$ and $\eta_r=0.1$.
In addition, a single-phase flow simulation has been also conducted.}
\label{overview}
\end{table}
\end{ruledtabular}

\section{Results}
\label{sec: results}

We present here the results obtained from the analysis of the simulation database, starting from the effects of the density ratio, viscosity ratio, and Weber number on the topology of the dispersed phase (number of bubbles) and on its topological changes (coalescence and breakage rates).
Then we evaluate the effects of these parameters on the shape and deformation of the bubbles studying the local curvature of the interface and the time evolution of the interfacial area.
Finally, we investigate the flow modifications produced by the bubbles by analyzing the mean velocity profiles and the turbulent kinetic energy inside the bubbles.
All the results will be presented according to the following color code: a red-colors scale is used to show the density ratio variations and a blue-colors scale to show the viscosity ratio variations. The case with both non-matched density and viscosity is represented in green, while the reference case (matched density and matched viscosity) is shown in black.

\subsection{Bubbles: number and topological modifications}


\subsubsection{Number of bubbles}

\begin{figure}
\setlength{\unitlength}{0.003\columnwidth}
\begin{picture}(400,370)
\put(45,163){\includegraphics[width=0.65\columnwidth, keepaspectratio]{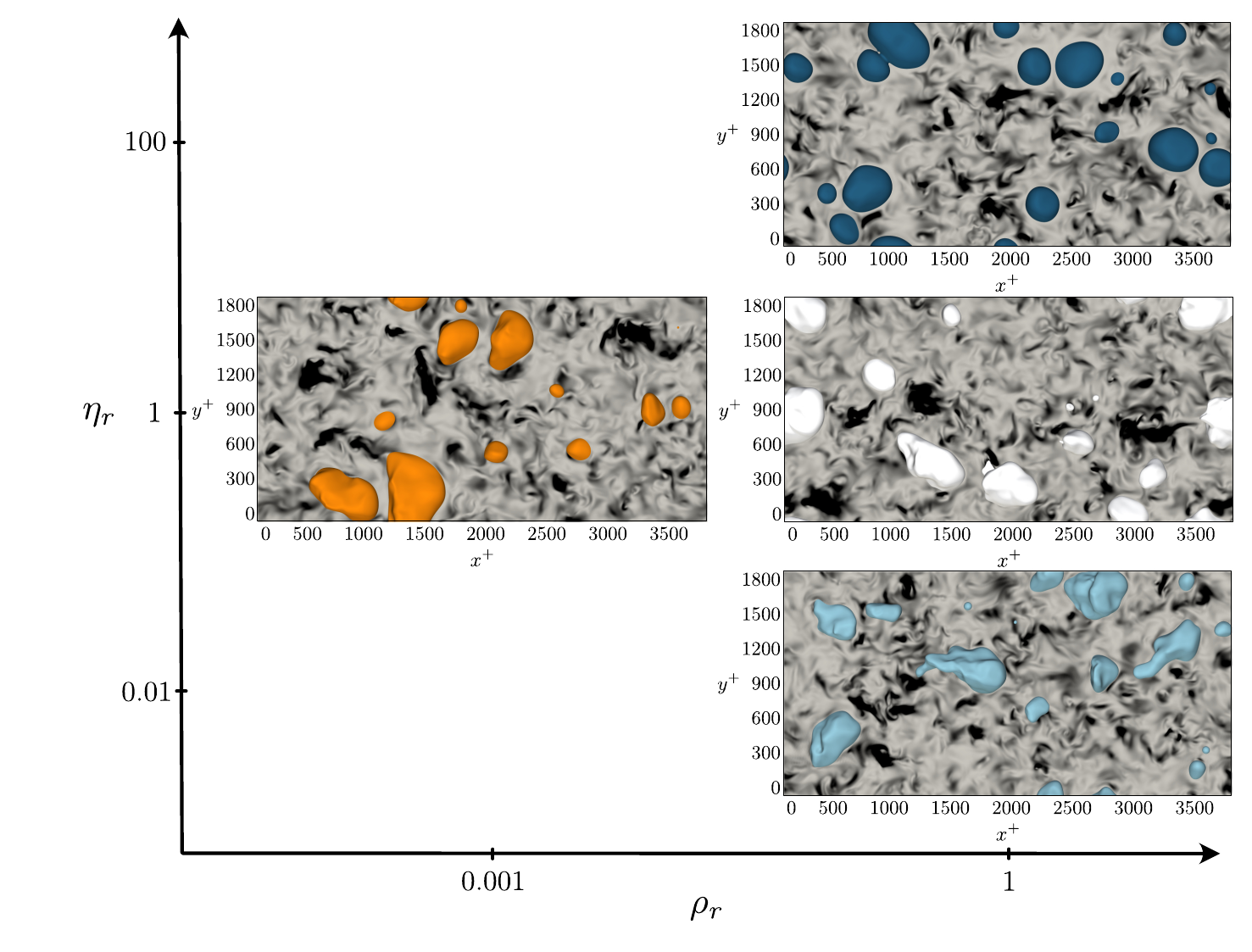}}
\put(45,-7){\includegraphics[width=0.65\columnwidth, keepaspectratio]{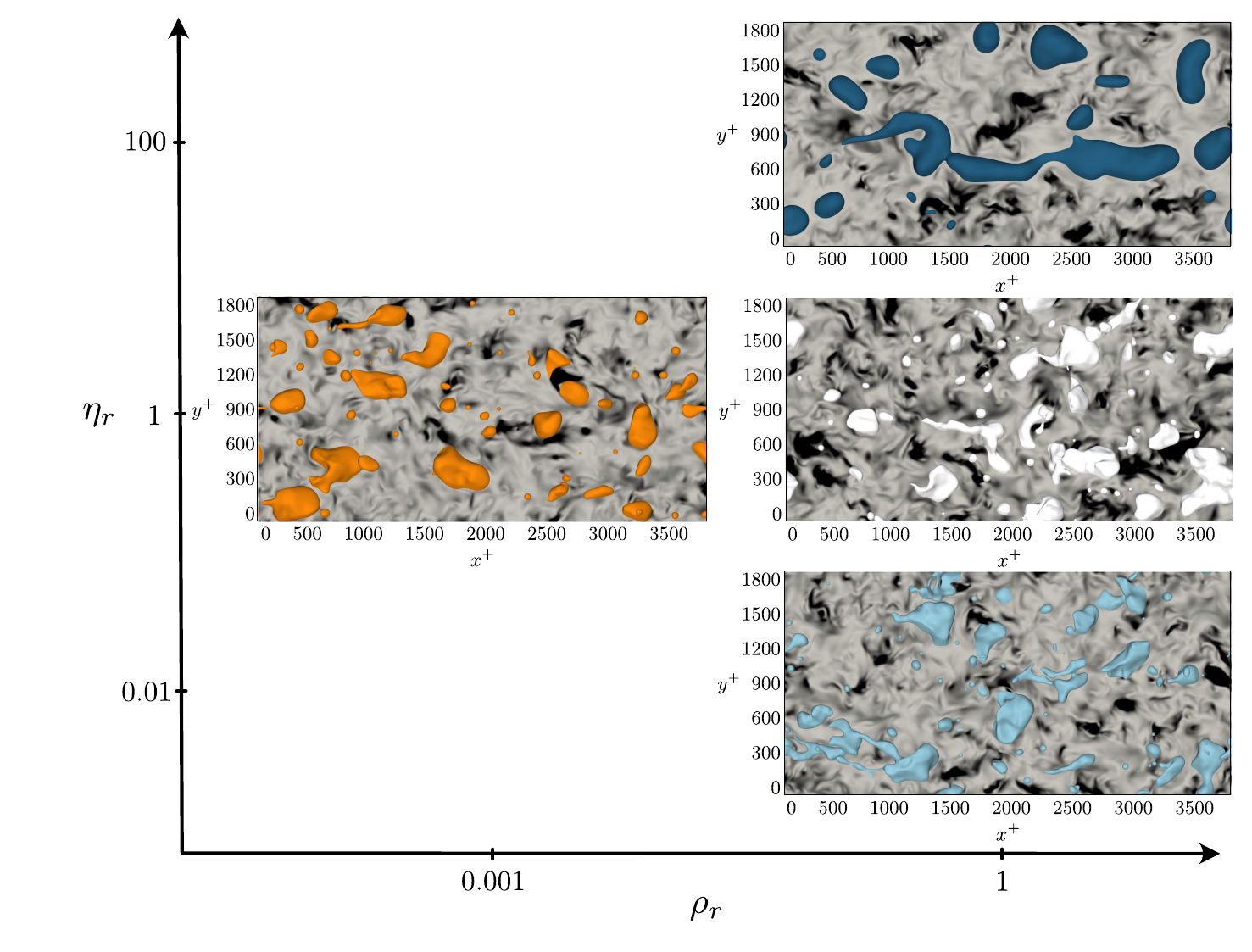}}
\put(230,61){\includegraphics[width=0.26\columnwidth, keepaspectratio]{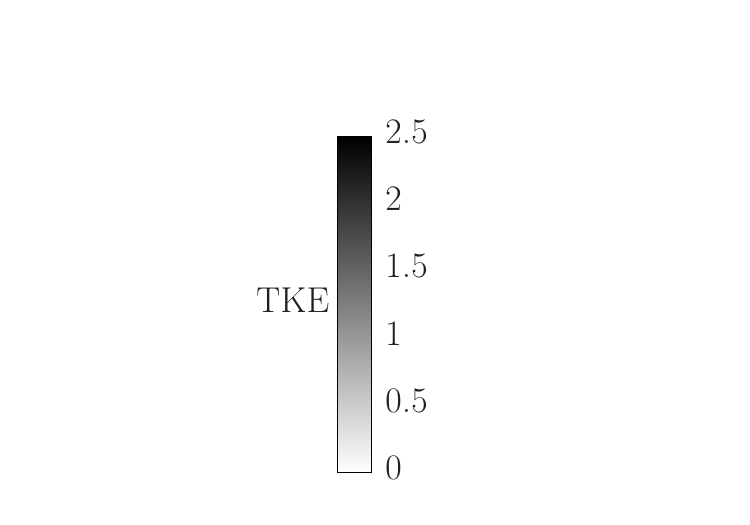}}
\put(230,231){\includegraphics[width=0.26\columnwidth, keepaspectratio]{1c}}
\put(60,315){(\textit{a})}
\put(60,145){(\textit{b})}
\put(110,300){$We=1.5$}
\put(110,130){$We=3.0$}
\end{picture}
\caption{Top view of four statistically-stationary configurations ($t^+=4000$) for different combinations of density ratios ($\rho_r= 0.001$ and $1$) and viscosity ratios ($\eta_r= 0.01,1$ and $100$).
Panel (\textit{a}) refers to $We=1.5$, while panel (\textit{b}) to $We=3.0$. 
The sub-panels are arranged in a plot using $\rho_r$ as $x$-coordinate and $\eta_r$ as $y$-coordinate. 
The effect of density can be appreciated in the sequence of panels on the middle row, while that of viscosity in the right column.
 The background of the plot shows the turbulent kinetic energy, TKE$=({\rho}/{\rho_c})(u'^2+v'^2+w'^2)/2$ (white-low; black-high), computed on the central $x^+ - y^+$ plane of the channel.}
 \label{fig:T}
\end{figure}

The topology of the dispersed phase is the direct consequence of the ultimate competition between breakage and coalescence events. 
To obtain a first qualitative insight of the effects of density ratio, viscosity ratio and Weber number on the statistically-stationary number of bubbles (i.e. once the effect of the initial condition is completely dissipated), we can consider figure~\ref{fig:T}.
Panel~(\textit{a}) refers to $We=1.5$, while panel~(\textit{b}) to $We=3.0$. 
In each panel of figure~\ref{fig:T}, four snapshots of the multiphase system at statistically-stationary are arranged in a plot according to the values of density (horizontal axis) and viscosity (vertical axis) ratio of each case.
The surface of the bubbles, identified as the iso-contour $\phi=0$,  is reported at the time instant $t^+=4000$ (statistically-stationary conditions); in the background the contour map of the turbulent kinetic energy, TKE$=({\rho}/{\rho_c})(u'^2+v'^2+w'^2)/2$ (where $\rho$ identifies the local density value, $\rho_d$ in the bubbles and $\rho_c$ in the carrier phase), on a $x^+-y^+$ plane located at the channel centre is shown.
Among all cases, we select those with the extreme values of the density ($\rho_r=0.001$ - $\eta_r=1$) and viscosity ratio ($\rho_r=1$ - $\eta_r=100$  and $\rho_r=1$ - $\eta_r=0.01$).
As a reference, also the matched density and viscosity case ($\rho_r=1$ - $\eta_r=1$) is shown.
We can observe that for $We=1.5$ (figure~\ref{fig:T}\textit{a}), the number of bubbles remains almost unchanged when both density and viscosity contrasts are introduced in the system. 
For $We=3.0$ (figure~\ref{fig:T}\textit{b}), the number of bubbles is higher in all the cases, compared to $We=1.5$. 
If we then look along the density axis (namely to the pictures in the central row) of figure~\ref{fig:T}\textit{b}, we see that the number of bubbles is quite similar in the two cases, suggesting a negligible effect of density for the range of values considered here.
By opposite, looking along the viscosity axis (thus to the pictures on the right column), we notice that viscosity does play an important role, as the number of bubbles significantly reduces from $\eta_r=0.01$ to $\eta_r=100$, with a more marked difference between $\eta_r=1$ and $\eta_r=100$, than between $\eta_r=0.01$ and $\eta_r=1$, thus hinting that the viscosity difference among the phases may actually be the relevant factor, rather than the viscosity ratio.

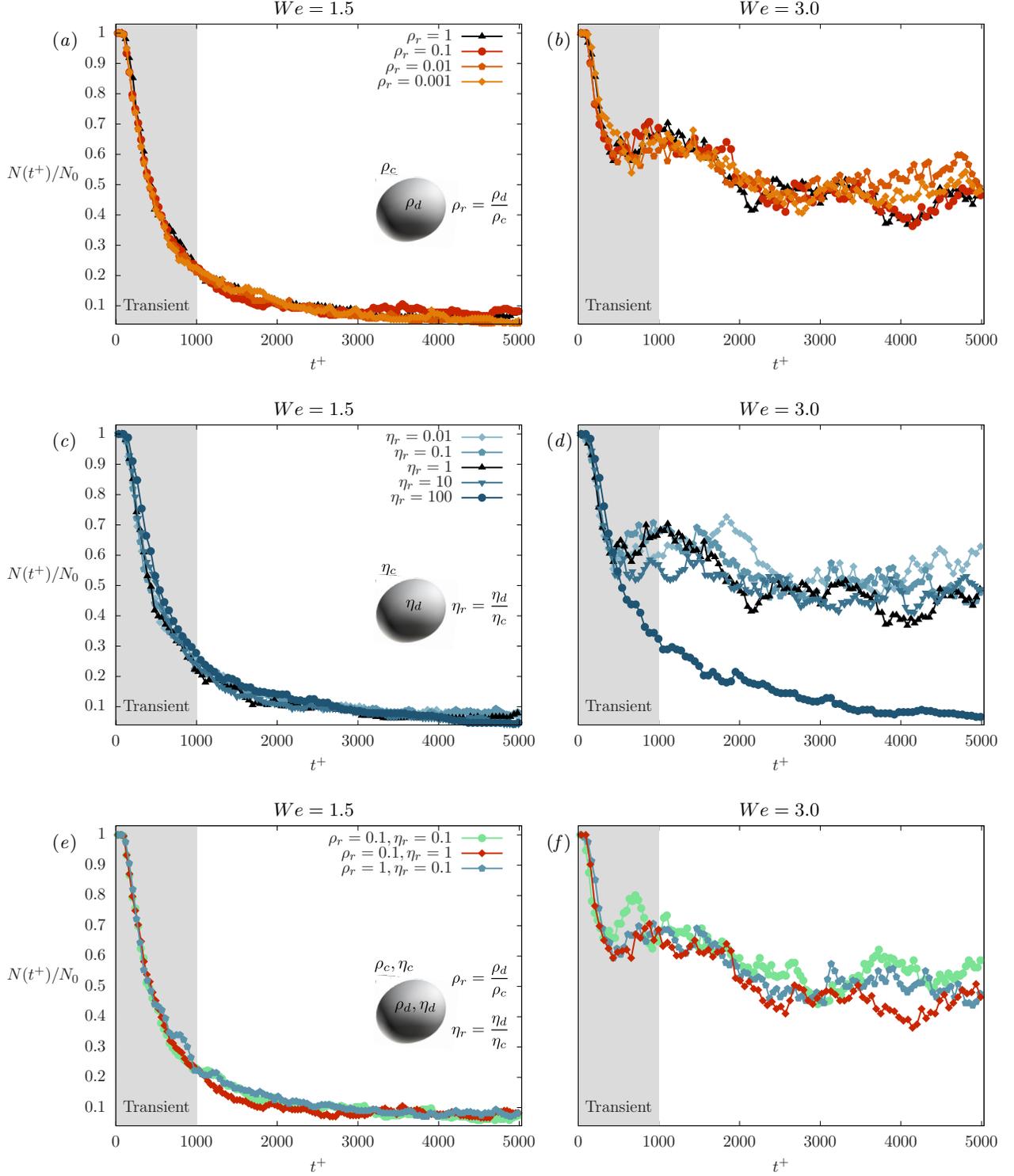
\begin{figure}
\setlength{\unitlength}{0.003\columnwidth}
\begin{picture}(400,375)
\put(238,365){$We=3.0$}
\put(93,365){$We=1.5$}
\put(10,250){\includegraphics[width=0.51\columnwidth, keepaspectratio]{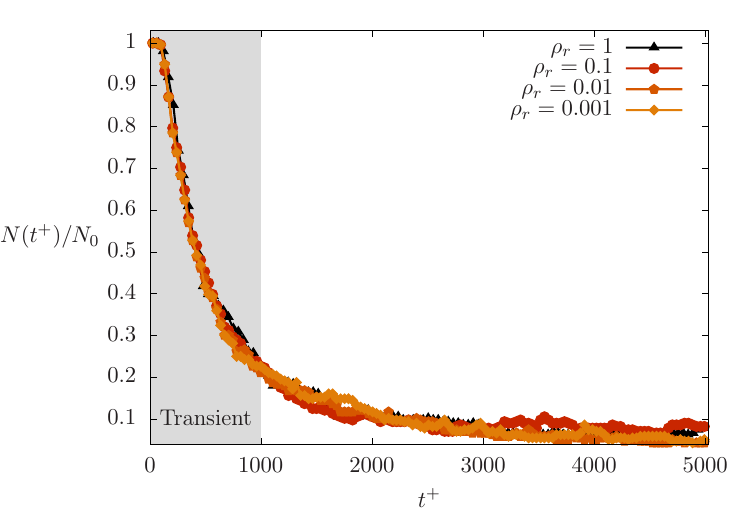}}
\put(154,250){\includegraphics[width=0.51\columnwidth, keepaspectratio]{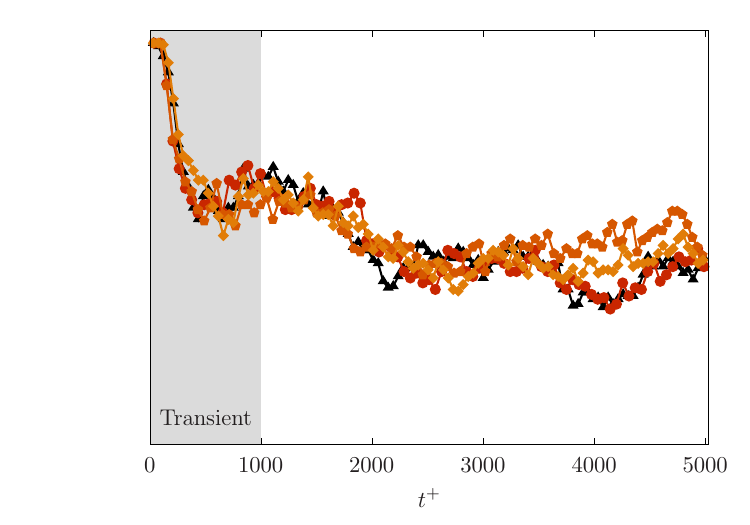}}
\put(124,290){\includegraphics[width=0.13\columnwidth, keepaspectratio]{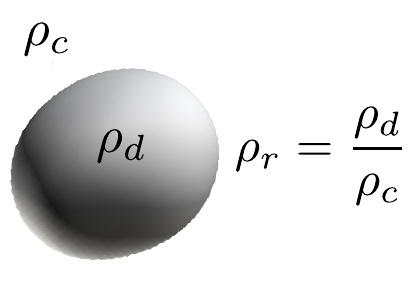}}
\put(24,355){(\textit{a})}
\put(178,355){(\textit{b})}
\put(238,240){$We=3.0$}
\put(93,240){$We=1.5$}
\put(10,125){\includegraphics[width=0.51\columnwidth, keepaspectratio]{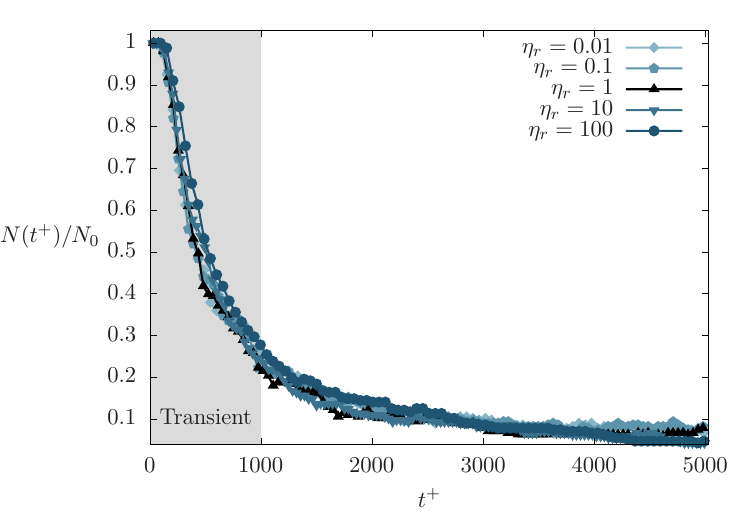}}
\put(154,125){\includegraphics[width=0.51\columnwidth, keepaspectratio]{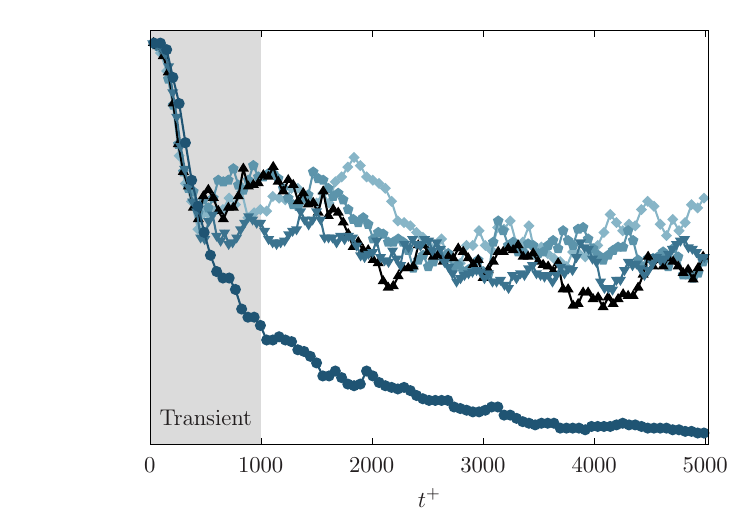}}
\put(124,165){\includegraphics[width=0.13\columnwidth, keepaspectratio]{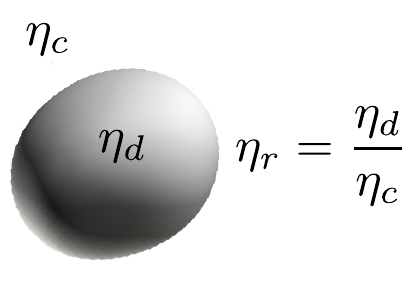}}
\put(24,230){(\textit{c})}
\put(178,230){(\textit{d})}
\put(238,115){$We=3.0$}
\put(93,115){$We=1.5$}
\put(10,0){\includegraphics[width=0.51\columnwidth, keepaspectratio]{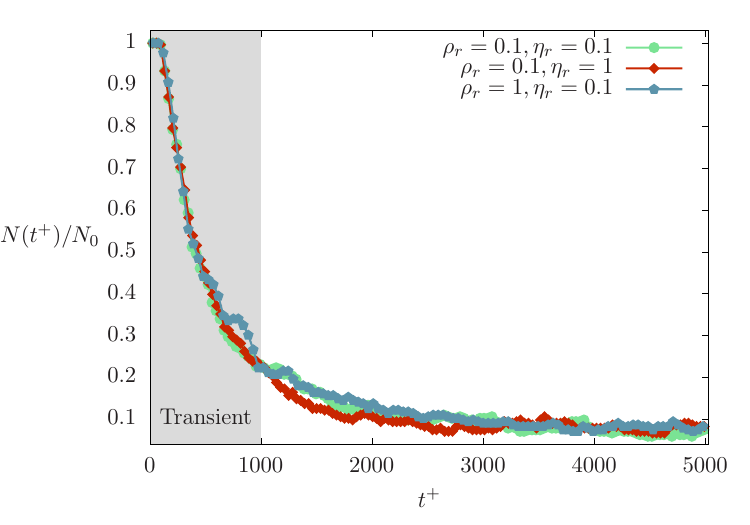}}
\put(154,0){\includegraphics[width=0.51\columnwidth, keepaspectratio]{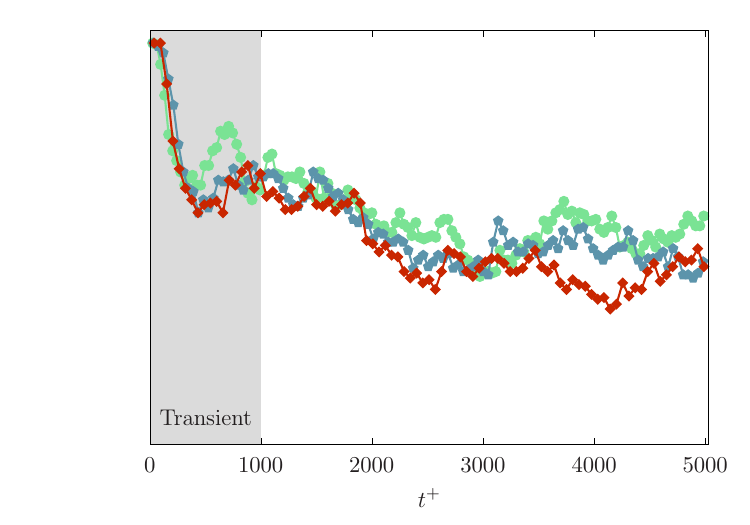}}
\put(124,40){\includegraphics[width=0.13\columnwidth, keepaspectratio]{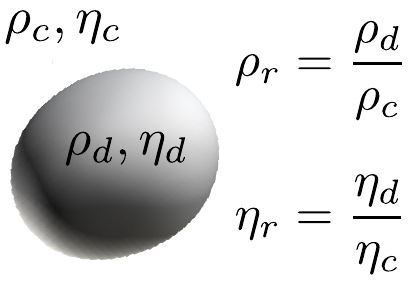}}
\put(24,105){(\textit{e})}
\put(178,105){(\textit{f})}
\end{picture}
\caption{Time evolution of the number of bubbles, $N(t^+)$, normalized by its initial value $N_0$. 
Left column refers to $We=1.5$, while the right column to $We=3.0$. 
Top row: effect of density ratio, for $\rho_r=0.001, 0.01, 0.1$ and $1$ (with $\eta_r=1$); Middle row: effect of viscosity ratio, for $\eta_r=0.01, 0.1, 1, 10$ and $100$ (with $\rho_r=1$); Bottom row: combined effect of density and viscosity, for the case with $\rho_r=0.1$, the cases $\rho_r=0.1$, $\eta_r=1$ and $\rho_r=1$, $\eta_r=0.1$ are reported for reference. 
On each line the left plot also includes the color code and a sketch with the definition of the property ratio considered ($\rho_r$, $\eta_r$ or both ratios).} 
\label{fig:DC}
\end{figure}

To evaluate these results more quantitatively, we compute at each time the number of bubbles, $N(t^+)$, normalized by the initial bubbles number, $N_0$. 
Figure~\ref{fig:DC} shows the results obtained for all the combination of density and viscosity ratios considered, and for the two Weber numbers as well. 
Left column refers to $We=1.5$ (figure~\ref{fig:DC}\textit{a},\textit{c},\textit{e}), while the right column to $We=3.0$ (figure~\ref{fig:DC}\textit{b},\textit{d},\textit{f}).
The top, middle and bottom rows show, in order, the effects of the density ratio, viscosity ratio and of their combination.

We start by analyzing the effect of Weber number solely and we consider the matched density and viscosity case (black lines in figure~\ref{fig:DC}\textit{a}-\textit{d}).
For $We=1.5$, the number of bubbles decreases monotonically: coalescence events dominate the initial transient phase (up to $t^+=2000$).
Then a balance between breakage and coalescence events is attained and the number of bubbles settles on a stationary value, $N(t^+)/N_0 \simeq 0.1$.
Likewise, for $We=3.0$, an initial transient mainly characterized by coalescence events can be also observed.
However, this phase ends at an earlier time (about $t^+=500$) and is followed by a statistically-stationary condition where breakups and coalescences alternately prevail on each other.

Comparing simultaneously the plots at $We=1.5$ (figure~\ref{fig:DC}\textit{a},\textit{c},\textit{e}), we can observe that the effects of both density and viscosity ratios (and of their combination) are very small.
This behavior can be traced back to the dominant role played by surface tension forces.
The Weber number quantifies the relative importance of surface tension forces with respect to inertial forces: the lower is the Weber number, the stronger is the action of surface tension in controlling bubbles dynamics.
Thus, for $We=1.5$, the surface tension forces are dominant and are those determining the topology of the dispersed phase (i.e. number of bubbles).
For the higher Weber, surface tension forces are weaker in comparison, and density and viscosity ratios effects become more significant. 
In particular, for $We=3.0$ (figure~\ref{fig:DC}\textit{b},\textit{d},\textit{f}), the statistically-stationary value obtained for the number of bubbles shows a marked dependence on the viscosity ratio.

As the dispersed phase dynamics for the cases at $We=1.5$ are dominated by surface tension forces, we focus on the cases at $We=3.0$ to investigate the effects of density and/or viscosity ratios.
First, we consider the effects of the density ratio solely.
Figure~\ref{fig:DC}\textit{b} shows the time evolution of the number of bubbles for different density ratios (from $\rho_r=1.0$ down to $\rho_r=0.001$) and a fixed unitary viscosity ratio.
We notice that the influence of the density ratio on the number of bubbles is small: the red-colors lines do not depart in average from the black reference line, nor from each other.
Hence, no significant modifications are introduced in the topology of the dispersed phase when density contrasts are present between the phases (with respect to a two-phase system with uniform density).
This behavior suggests that, for the range of density ratios considered, the external inertial forcing is the main factor that determines the bubble size and thus the dispersed phase topology.
In contrast, the density (and thus the inertia) of the bubble plays a negligible role in determining  the dispersed phase topology.

On the other hand a marked effect of the viscosity ratio alone can be observed, figure~\ref{fig:DC}\textit{d}.
We observe in this case a much clearer trend: after the initial transient the curves depart from each other and set on different equilibrium values once statistically-stationary conditions are reached.
In particular, as the viscosity ratio is increased, the statistically-stationary number of bubbles is reduced. 
For high viscosity ratio ($\eta_r>1$) fragmentation is prevented, coalescence dominates and only a few bubbles are present in the channel.
By opposite, for low viscosity ratio ($\eta_r<1$) breakups are favored, the average bubble size decreases, and the resulting number of bubbles is slightly larger when smaller viscosity ratios are considered.
Hence, it is evident that viscosity acts as a stabilizing factor, in a similar way as surface tension does. 
Indeed, it is interesting to observe that the behavior of the number of bubbles for $\eta_r=100$ at $We=3.0$ (high viscosity) resembles those of the cases at $We=1.5$ (high surface tension, figure~\ref{fig:DC}\textit{c}).
This suggests that a very high viscosity ratio can compensate a low surface tension and produce similar results in terms of topology. 
A physical argument that can explain the action of viscosity is related to the deformations that the external turbulent flow is able to induce on the bubble. 
When the internal viscosity is larger than the external one, the larger internal viscous dissipation damps all the turbulent fluctuations produced by the external flow.
This hinders large deformations of the bubble surface and, as a consequence, it reduces the possibility of bubble breakage.


Finally, we analyze the combined effects of density and viscosity ratios.
In figure~\ref{fig:DC}\textit{f}, we report the results obtained from the case $\rho_r=0.1$ and $\eta_r=0.1$ and from two cases with one matched property and one non-matched property, $\rho_r=0.1$ and $\eta_r=1$ (red line) and $\rho_r=1$ and $\eta_r=0.1$ (blue line).
We can first note that these two latter cases, where only one property is non-matched, exhibit a very similar behavior for the entire duration of the simulation. 
This is consistent with our previous observation: the influence of the density ratio is almost negligible (figure~\ref{fig:DC}\textit{b}) and the effects of the viscosity ratio are relatively small for $\eta_r = 0.1$ (figure~\ref{fig:DC}\textit{d}).
Then, we observe that the combined case (green line) does not deviate largely from the other two cases.
This indicates that a simultaneous reduction of the density and viscosity ratios does not remarkably modify the general picture for the range of density and viscosity ratios here tested.
Nevertheless, it is interesting to observe that the green line lies above the red and blue lines for a longer timespan, indicating that the number of bubbles for the combined case is slightly higher than in the other two cases.

\subsubsection{Breakage and coalescence rates}
 \label{sssec:bcrate}
 
The evolution of the number of bubbles provides useful insights on the time behavior of the dispersed phase topology, although it only shows the net outcome of the competition between breakage and coalescence events. 
To evaluate whether density and viscosity differences among the phases affect breakage and coalescence dynamics, we compute the instantaneous number of breakage and coalescence events.
Evaluating these effects is not only crucial to better understand the involved physics, but is also extremely important for the development of accurate coalescence and breakage kernels \cite{Falzone2018}.
The time behavior of the breakage and coalescence is directly linked to the number of bubbles present in the channel,  as hinted by the balance population equation \cite{lasheras2002review}:
\begin{equation}
\frac{dN(t^+)}{dt^+}=\dot{N_b}(t^+) - \dot{N_c}(t^+) \; ,
\end{equation}
where $N(t^+)$ is the number of bubbles and $\dot{N_b}(t^+)$ and $\dot{N_c}(t^+)$ are respectively the breakage and coalescence rates.
We compute the breakage and coalescence rates counting the number of breakage or coalescence events that occur within a set temporal window $\Delta t^+$ (see Appendix~\ref{appa} for details):
\begin{equation}
\dot{N_b}(t^+)=\frac{N_b}{\Delta t^+},  \hspace{1cm} \dot{N_c}(t^+)=\frac{N_c}{\Delta t^+}\, ,
\end{equation}
where the temporal window has been chosen equal to $\Delta t^+=300$.
As the number of breakage and coalescence events that occur in a certain temporal window is also influenced by the number of bubbles present in the channel \cite{Soligo2019c}, we normalize the breakage and coalescence rates, $\dot{N_b}(t^+)$ and $\dot{N_c}(t^+)$, by the instantaneous number of bubbles $N(t^+)$. 
Being the description of coalescence and breakage events in numerical simulations influenced by grid resolution \cite{gorokhovski2008modeling,tryggvason2013multiscale,Soligo2019c,soligo2021}, a convergence study has been also performed to ensure that the grid employed is sufficient to obtain convergent results, please refer to Appendix~\ref{appb} for details.

\begin{figure}
\setlength{\unitlength}{0.003\columnwidth}
\begin{picture}(400,375)
\put(238,365){$We=3.0$}
\put(93,365){$We=1.5$}
\put(10,250){\includegraphics[width=0.51\columnwidth, keepaspectratio]{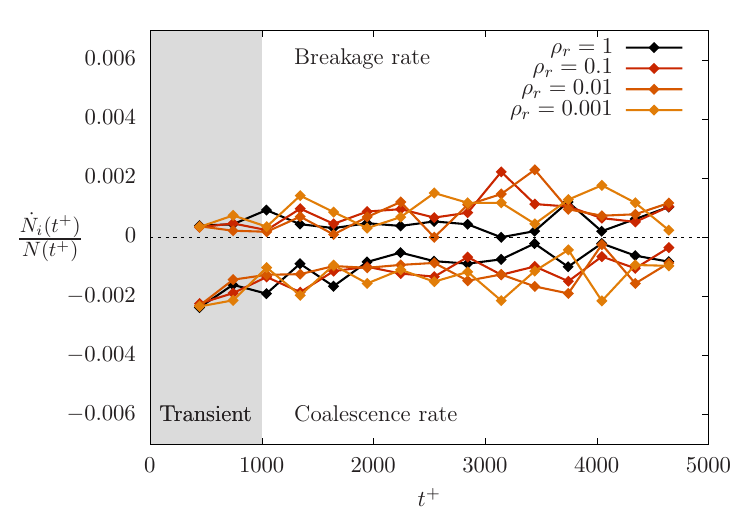}}
\put(154,250){\includegraphics[width=0.51\columnwidth, keepaspectratio]{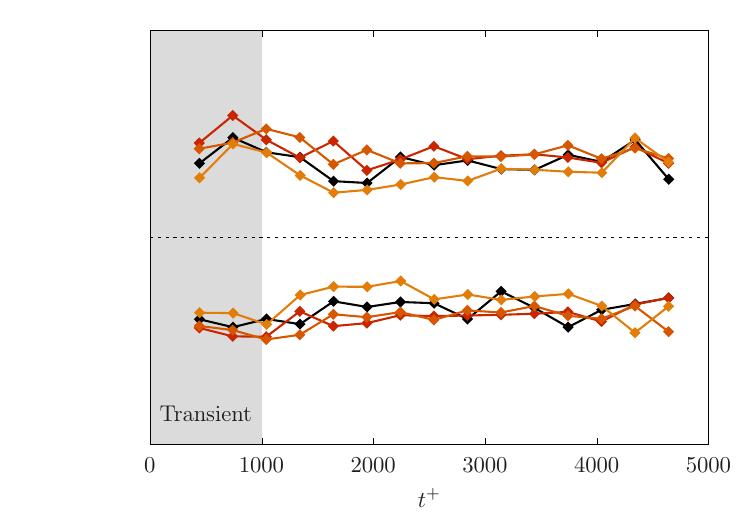}}
\put(124,268){\includegraphics[width=0.13\columnwidth, keepaspectratio]{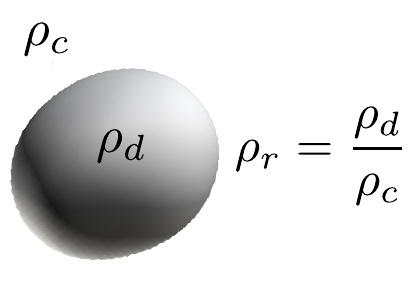}}
\put(14,355){(\textit{a})}
\put(178,355){(\textit{b})}
\put(238,240){$We=3.0$}
\put(93,240){$We=1.5$}
\put(8,125){\includegraphics[width=0.51\columnwidth, keepaspectratio]{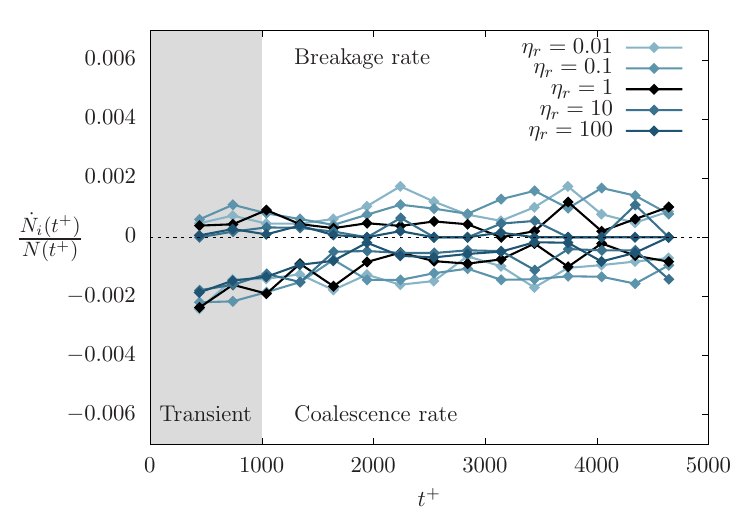}}
\put(154,125){\includegraphics[width=0.51\columnwidth, keepaspectratio]{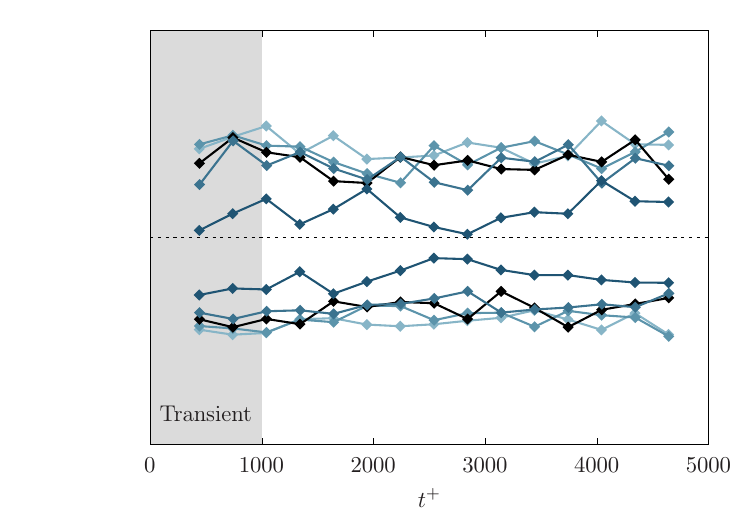}}
\put(124,143){\includegraphics[width=0.13\columnwidth, keepaspectratio]{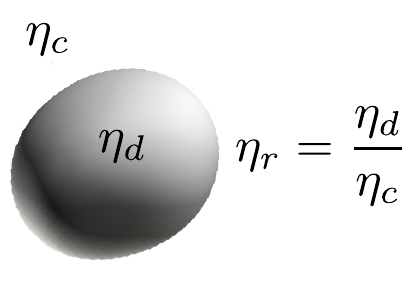}}
\put(14,230){(\textit{c})}
\put(178,230){(\textit{d})}
\put(238,115){$We=3.0$}
\put(93,115){$We=1.5$}
\put(8,0){\includegraphics[width=0.51\columnwidth, keepaspectratio]{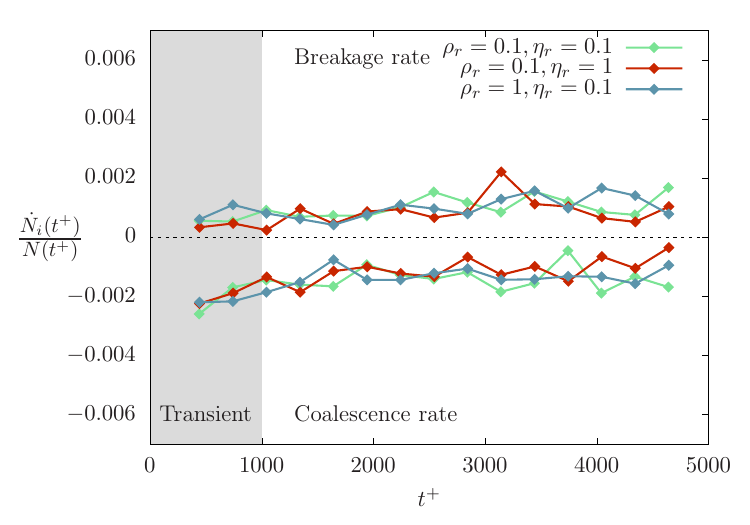}}
\put(154,0){\includegraphics[width=0.51\columnwidth, keepaspectratio]{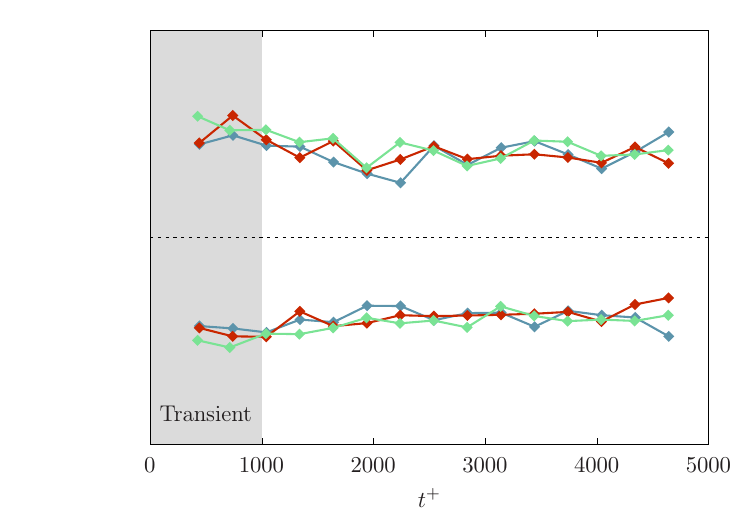}}
\put(124,18){\includegraphics[width=0.13\columnwidth, keepaspectratio]{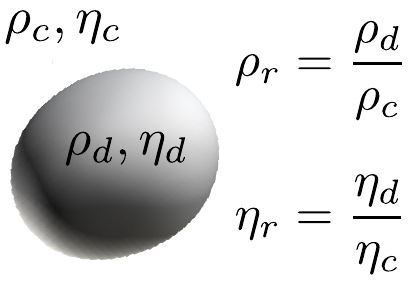}}
\put(14,105){(\textit{e})}
\put(178,105){(\textit{f})}
\put(238,115){$We=3.0$}
\put(93,115){$We=1.5$}
\end{picture}
\caption{Time evolution of the normalized breakage rate, $\dot{N}_b(t^+)/N(t^+)$, and coalescence rate, $\dot{N}_c(t^+)/N(t^+)$. 
Left column refers to $We=1.5$, while right column to $We=3.0$.
Top row: effect of density ratio, for $\rho_r=0.001, 0.01, 0.1$ and $1$ (with $\eta_r=1$); Middle row: effect of viscosity ratio, for $\eta_r=0.01, 0.1, 1, 10$ and $100$ (with $\rho_r=1$);  Bottom row: combined effect of density and viscosity ratios, for the case with $\rho_r=0.1$, $\eta_r=0.1$. 
Cases  $\rho_r=0.1$, $\eta_r=1$ and $\rho_r=1$, $\eta_r=0.1$ are reported for reference.
For each row of plots, the left plot also shows the color code and a sketch with the definition of the ratio considered ($\rho_r$, $\eta_r$ or both).} 
\label{fig:CBn}
\end{figure}
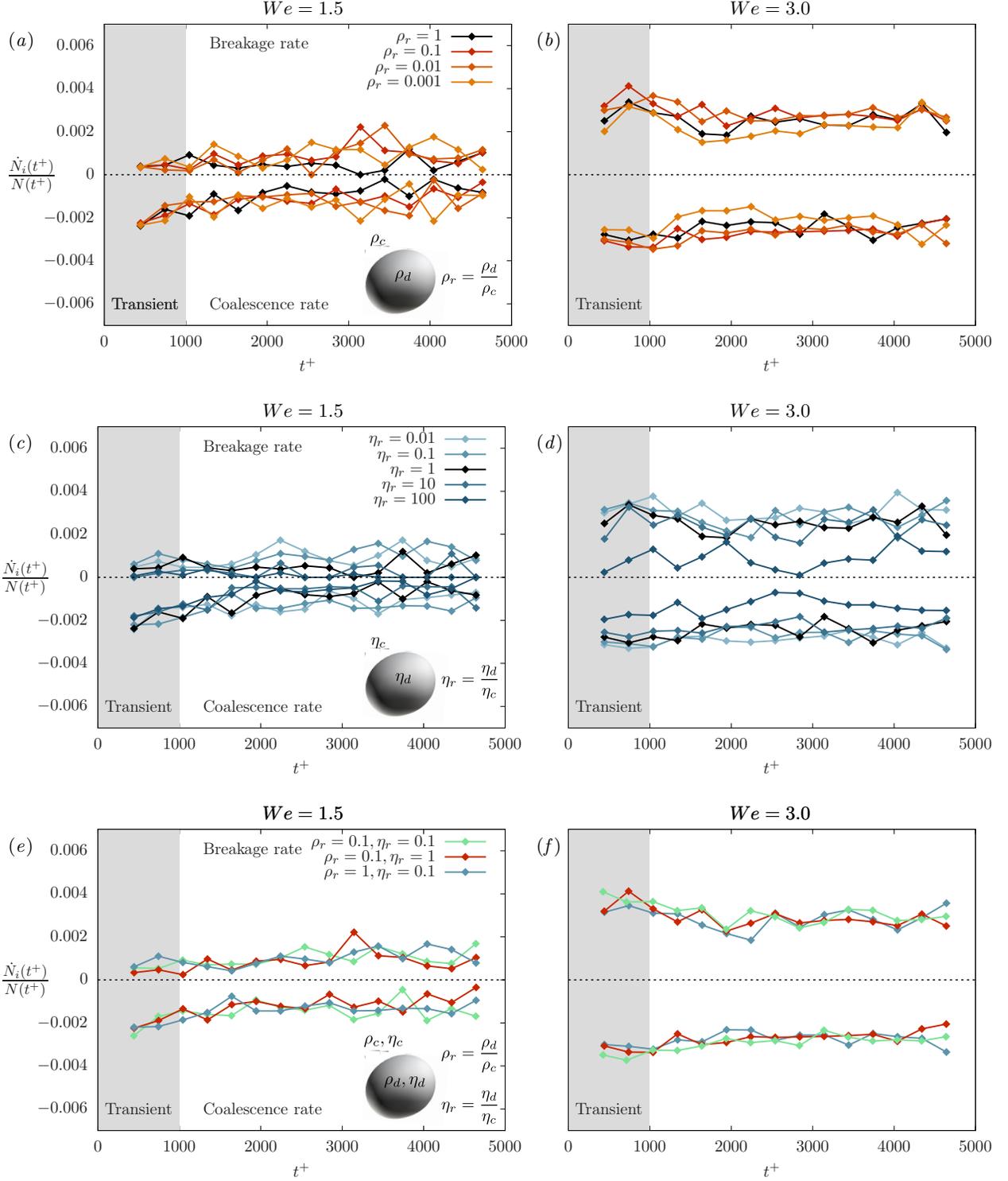

Figure~\ref{fig:CBn} shows the results obtained for all cases examined: breakage rate is plotted over time as a positive quantity, while coalescence rate as a negative quantity, being them related to an increase and decrease of the number of bubbles, respectively. 
We will first discuss the effect of the Weber number comparing the left column (figure~\ref{fig:CBn}\textit{a},\textit{c},\textit{e}) with the right column (figure~\ref{fig:CBn}\textit{b},\textit{d},\textit{f}).
For $We=1.5$ (left column), the breakage and coalescence rates behave nearly in the same way for all the combinations of density and viscosity ratios.
After the initial transient where the behavior of the rates is influenced by the selected initial condition for the phase-field, an equilibrium is reached at about $t^+=1000$ where both rates set on a constant value.
At this stage, bubbles keep on breaking and coalescing, but with the same rate, thus maintaining their number in statistical equilibrium. 
This value of the Weber number does not allow density and viscosity contrasts to substantially modify the evolution of bubbles topology, as a good correspondence among the curves can be noticed in all the plots.
Indeed, when a low Weber number is considered the deformability, which is a crucial factor for coalescence and breakage events, is mainly determined by surface tension forces that dominate over density and viscosity contributions. 
For $We=3.0$ (right column), the results are qualitatively and quantitatively different: breakage and coalescence rates reach in general larger values, and some significant deviations among the curves are visible. 
This is a direct consequence of the larger Weber number: surface tension forces, which are smaller in magnitude, weakly counteract turbulent velocity gradients, that can more easily deform and break the bubbles. 
Thus, we observe a larger number of breakage and coalescence events due to the larger deformability of the bubbles, as can be appreciated from figure~\ref{fig:DC}\textit{b},\textit{d},\textit{f}.
In addition, for this larger Weber number, we can clearly observe how the density and viscosity ratios play a much more important role in the dynamics of breakage and coalescence events (with respect to $We=1.5$).

For this reason, we move now to discuss the effect of non-matched density or viscosity on the cases at $We=3.0$ in more detail.
Figure~\ref{fig:CBn}\textit{b} shows the breakage and coalescence rate for different values of the density ratios. 
In the first transient phase, all cases manifest a very high frequency of both breakage and coalescence events, slightly larger for coalescences at the very beginning (coherently with the evolution of the number of bubbles shown in figure~\ref{fig:DC}\textit{b}).
Later on, both rates stabilize and set on two equal (in magnitude) stationary values. 
Although a clear trend among the different density ratios cannot be observed, it is worth noticing that all the rates seem slightly larger when sub-unitary density ratios are considered (especially in the early stage of simulations).
Overall, these observations suggest that density differences between the phases do not introduce remarkable changes in the dispersed phase topology and on its modifications: the number of bubbles and breakage and coalescence rates are weakly influenced by changing the density ratio.

Moving now to the effect of the viscosity ratio, figure~\ref{fig:CBn}\textit{d} depicts the time evolution of the breakage and coalescence rates obtained for different viscosity ratios (and a fixed unitary density ratio).
Again, once the initial transient is finished, a statistically-stationary phase can be distinguished for all cases.
From a qualitative viewpoint, coalescence is predominant at the beginning of the transient (consistently with the behavior reported in figure~\ref{fig:DC}\textit{d}); then relatively high values for both rates are maintained during the rest of the transient, until they stabilize on steady values. 
The cases, however, deeply differ from a quantitative point of view. 
We see in this case that the rates significantly change when the viscosity ratio is changed: both breakage and coalescence rates decrease in magnitude as the viscosity ratio is increased (i.e. when bubble viscosity is increased). 
This modification of the breakage and coalescence rates is clear when the case $\eta_r=100$ is considered: the statistically-steady value of both rates is smaller than the one attained by the other cases. 
A similar trend was experimentally measured by \citet{eastwood2004breakup} for the breakup of immiscible fluid particles in a turbulent jet: it was observed that the breakage rate of the droplets scales inversely with the inner bubble capillary number (ratio between bubble viscous forces and surface tension forces).
Present results seem to confirm this finding: bubble viscosity and the corresponding viscous forces, acting as a damper of external velocity fluctuations \cite{Roccon2017}, make bubbles less deformable and the probability of breakage and coalescence decreases.

Finally, we discuss the combination of density and viscosity contrasts (figure~\ref{fig:CBn}\textit{f}). 
The three curves do not deviate considerably from each other and a clearcut trend cannot be appreciated.
As the density effect is generally unimportant and the viscosity one shall be small for $\eta_r=0.1$, the case $\rho_r=0.1$ - $\eta_r=0.1$ does not give us clear information on how density and viscosity effects combine together.

\subsection{Shape and deformation of bubbles}

\subsubsection{Interfacial area} 

A bubble released in a turbulent flow is constantly subjected to deformations due to the action of turbulent fluctuations \cite{Perrard_2021,velamartin2021}. 
Turbulence fluctuations deform and stretch the bubble and, if strong enough, can lead to  breakage of the bubble.
The result of turbulence actions in terms of deformation can be evaluated by computing the total interfacial area.
This quantity gives a general indication of the average bubble deformation and  also provides a quantification of the amount of energy stored at the interface \cite{joseph1976stability,Dodd2016,roccon2021}.
Indeed, in the hypothesis of constant surface tension (as in the present case), surface tension energy is proportional to the amount of interfacial area available \cite{joseph1976stability,Dodd2016,roccon2021}.

\begin{figure}
\setlength{\unitlength}{0.003\columnwidth}
\begin{picture}(400,370)
\put(238,365){$We=3.0$}
\put(93,365){$We=1.5$}
\put(10,250){\includegraphics[width=0.51\columnwidth, keepaspectratio]{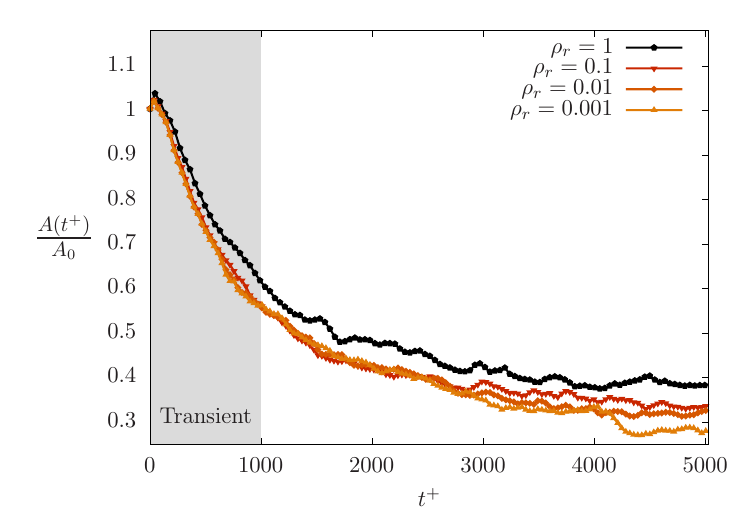}}
\put(154,250){\includegraphics[width=0.51\columnwidth, keepaspectratio]{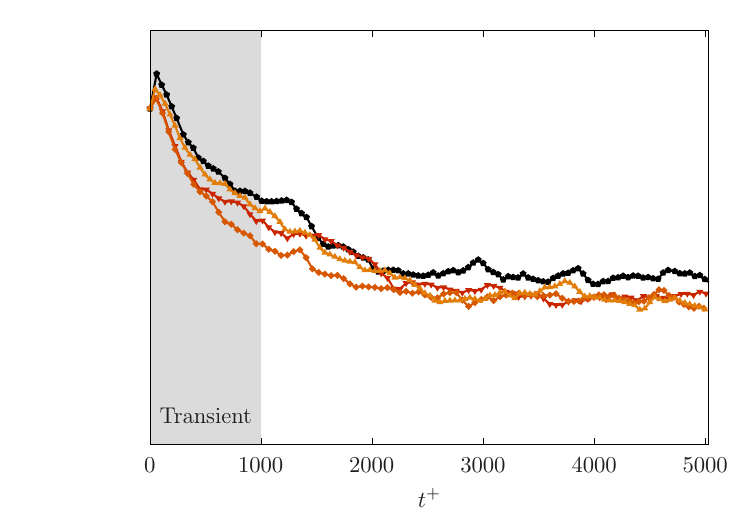}}
\put(124,294){\includegraphics[width=0.13\columnwidth, keepaspectratio]{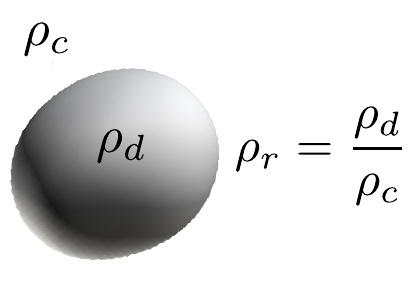}}
\put(24,355){(\textit{a})}
\put(178,355){(\textit{b})}
\put(238,240){$We=3.0$}
\put(93,240){$We=1.5$}
\put(10,125){\includegraphics[width=0.51\columnwidth, keepaspectratio]{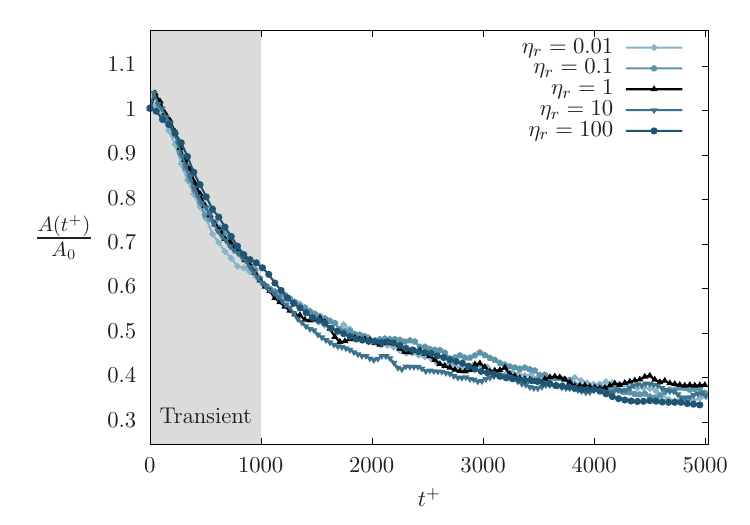}}
\put(154,125){\includegraphics[width=0.51\columnwidth, keepaspectratio]{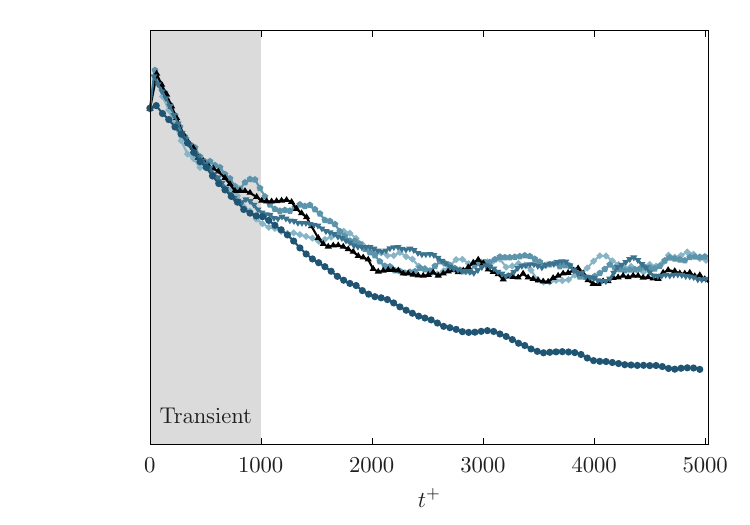}}
\put(124,172){\includegraphics[width=0.13\columnwidth, keepaspectratio]{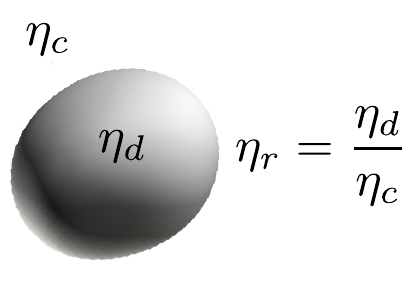}}
\put(24,230){(\textit{c})}
\put(178,230){(\textit{d})}
\put(238,115){$We=3.0$}
\put(93,115){$We=1.5$}
\put(10,0){\includegraphics[width=0.51\columnwidth, keepaspectratio]{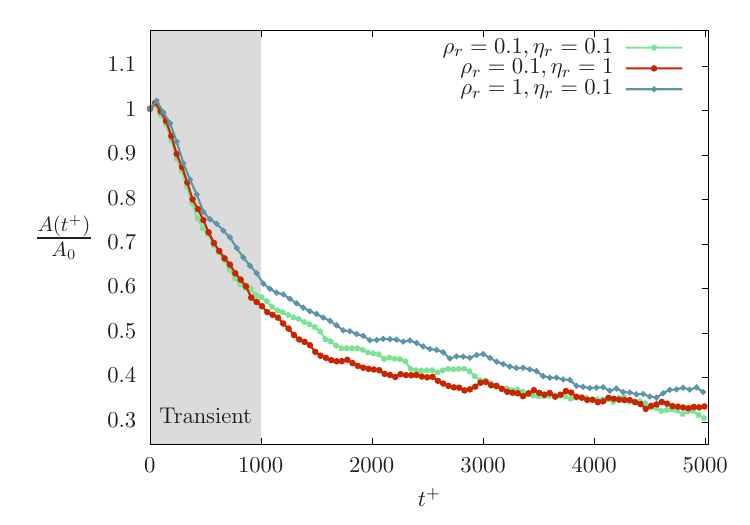}}
\put(154,0){\includegraphics[width=0.51\columnwidth, keepaspectratio]{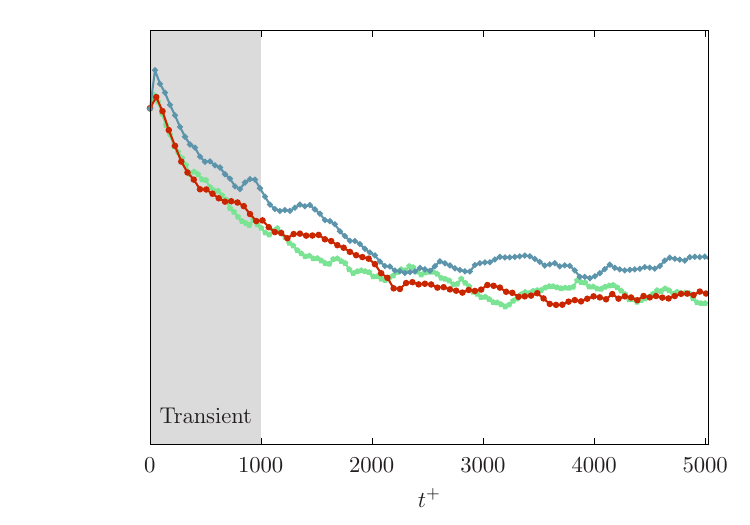}}
\put(124,45){\includegraphics[width=0.13\columnwidth, keepaspectratio]{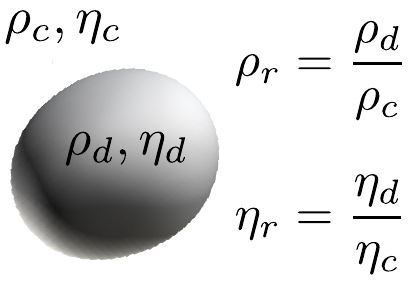}}
\put(23,105){(\textit{e})}
\put(178,105){(\textit{f})}
\end{picture}
\caption{Time evolution of the total interface area $A(t^+)$, normalized by its initial value $A_0$. 
Top row: effect of density, for $\rho_r=0.001, 0.01, 0.1$ and $1$ (with $\eta_r=1$); Middle row: effect of viscosity, for $\eta_r=0.01, 0.1, 1, 10$ and $100$ (with $\rho_r=1$); Bottom row: combined effect of density and viscosity, for the case with $\rho_r=0.1$, $\eta_r=0.1$. Cases with $\rho_r=0.1$, $\eta_r=1$ and $\rho_r=1$, $\eta_r=0.1$ are reported for reference. 
These effects are shown for two different Weber numbers: (\textit{a})-(\textit{c})-(\textit{e}) $We=1.5$ and (\textit{b})-(\textit{d})-(\textit{f}) $We=3.0$. On each row the left plot also includes the colour code and a sketch with the definition of the ratio considered ($\rho_r$, $\eta_r$ or both ratios).} 
\label{fig:AI}
\end{figure}
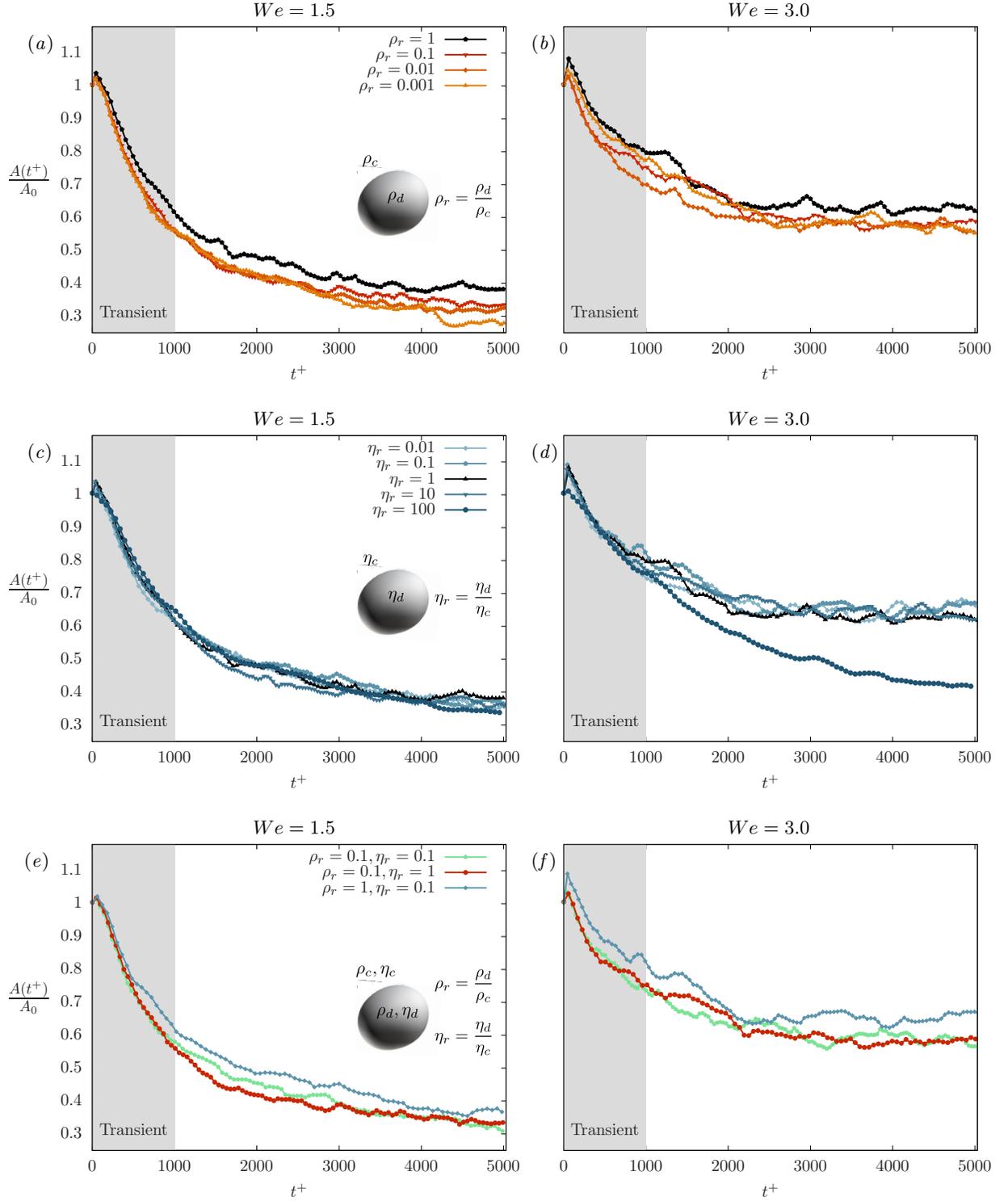

With the aim of evaluating the effects of the simulations parameters (density ratio, viscosity ratio, and Weber number) on the interfacial energy, we compute the time behavior of total interfacial area, $A(t^+)$, for all cases considered.
The results are presented normalized by the initial value $A_0$.
In figure~\ref{fig:AI}, the results are shown using the same arrangement adopted in the previous figures.
To correctly interpret these results, it is necessary to make a preliminary remark. 
The area of the interface between the dispersed phase and the carrier fluid evolves in time depending on two factors: the evolution of the number of bubbles and the modifications of the shape of the bubbles. 
This concept can be explained by considering the following example: to have a minimal interface area, the dispersed phase should consist of a unique spherical bubble, since, for a given volume, the spherical shape is the one that minimizes the surface area. 
If we split this bubble into several smaller spherical bubbles the total interface area will increase, being the total volume constant. 
If these smaller spherical bubbles are then deformed and elongated the area will further increase, as for each bubble the same mass will be redistributed in a way that makes it more exposed to the external flow. 
Thus, when we look at the evolution of the total interface area we are simultaneously observing the effect of the number of bubbles and of their deformation.
We start by analyzing the effects of the density ratio for the cases at $We=1.5$, figure~\ref{fig:AI}\textit{a}.
We notice an initial transient that is characterized by a nearly monotonic decrease of $A(t^+)/A_0$, for all the considered cases. 
In particular, during this transient, the curves corresponding to sub-unitary density ratios are superposed, while a remarkable discrepancy is visible between them and $\rho_r=1$. 
As soon as the flow reaches a steady behavior, all the curves differentiate and a trend becomes visible, where the higher is the density ratio the larger is the total interface area. 
Considering that for $We=1.5$ the number of bubbles is almost unaffected by the density ratio (figure~\ref{fig:DC}\textit{a}), this indicates that the trends observed in figure~\ref{fig:AI}\textit{a} are mainly caused by the bubble deformation: when smaller density ratios are considered, bubbles tend to be less deformed with respect to the case $\rho_r=1$.
The origin of this behavior can be traced back to the local Reynolds number (i.e. evaluated using the bubble proprieties): as the density ratio is decreased, the inertial forces become smaller, the local Reynolds number decreases and less deformed bubbles are obtained.

For $We=3.0$ (figure~\ref{fig:AI}\textit{b}), we notice a similar but more irregular behavior. 
For all density ratios, the normalized interfacial area decreases and sets on stationary values that are higher than the final stationary values obtained for $We=1.5$ (figure~\ref{fig:AI}\textit{a}). 
This is coherent with the fact that increasing the Weber number, the number of bubbles increases, and so does the interfacial area. 
For this larger Weber number, the trend among the different density ratios is now less clear and the differences between the curves are slightly smaller.
Nevertheless, consistently with the results obtained for $We=1.5$ (figure~\ref{fig:AI}\textit{a}), the matched density case ($\rho_r=1$) is clearly above all the other curves ($\rho_r<1$) for almost the entire time range of the simulations.
Being the number of bubbles similar for all the cases shown in figure~\ref{fig:AI}\textit{b}, this seems to confirm that for smaller density ratios the overall interfacial area is reduced.

The viscosity effect can be appreciated in figure~\ref{fig:AI}\textit{c},\textit{d}.
For $We=1.5$ (panel \textit{c}), the total interface area is practically independent on the viscosity ratio and no significant changes can be observed.
As the number of bubbles is similar for all cases (figure~\ref{fig:DC}\textit{c}), this indicates that no significant effects on the average bubble deformation are observed.
Even though bubble viscosity does not play an important role in the average bubble deformation, we can anticipate that it still plays a role when more local quantities are analyzed (e.g. local curvature), see  section~\ref{sec: curvature}.
For $We=3.0$, a remarkable difference is present between $\eta_r=100$ (larger bubble viscosity) and all the other cases. 
This is consistent with the time evolution of the number of bubbles (figure~\ref{fig:DC}\textit{d}).
Indeed, when the statistically-stationary configuration is reached, the number of bubbles for $\eta_r=100$ is much lower than that obtained for the other ratios.
As a result, the interfacial area is much lower than the other cases.
For the other cases (from $\eta_r=10$ down to $\eta_r=0.01$), a clear trend cannot be observed thus suggesting that no large modifications of the average bubble deformation are obtained for $\eta_r <10$. 
However, as already anticipated for $We=1.5$, larger modifications are observed when local quantities are analyzed, see next section for details.

Finally, we discuss the combined effect of density and viscosity ratios (figure~\ref{fig:AI}\textit{e},\textit{f}). 
For $We=1.5$, the case with both non-matched density and viscosity (green line) overlaps the case with non-matched density (red line) during the transient and in the final steady configuration, while in the first steady part it is intermediate between the two other cases, $\rho_r=0.1$ - $\eta_r=1$ and $\rho_r=1$ - $\eta_r=0.1$. 
On average the combined case is therefore closer to the non-matched density case, suggesting that the density ratio has a larger influence on the total interfacial area (and thus on the stretching of the bubbles) with respect to the viscosity ratio.
This is confirmed by the plot for $We=3.0$, where the green line shows again values that on average are much closer to the non-matched density case (i.e. $\rho_r=0.1$).

\subsubsection{Probability density function of mean curvature}
\label{sec: curvature}

The evolution of the total interface area gives us an idea of the overall behavior of the average deformation of the bubbles in presence of density and viscosity contrasts. 
However, being an average indication, it does not provide a clear indication of the local deformations of the bubbles surface.
To obtain a deeper understanding of the deformation, we examine the probability density function (PDF) of the local interface mean curvature in the final statistically-stationary configuration. 
The mean curvature, $\mathcal{K}^+$, can be computed as the divergence of the local normal vector \textbf{n}, which in turn can be defined from the phase variable $\phi$ \cite{Aris1989,Sun2007}:
\begin{equation}
\textbf{n}=-\frac{\nabla \phi}{\lvert \nabla \phi \rvert}, \hspace{1cm} \mathcal{K}^+=\nabla \cdot \bigg(-\frac{\nabla \phi}{\lvert \nabla \phi \rvert}\bigg).
\end{equation} 
We compute the mean curvature, $\mathcal{K}^+$, for each point on the surface of the bubbles, corresponding to the points of the iso-level $\phi=0$. 
The resulting curvature values tell us how much the bubbles deviate from their spherical equilibrium shape, giving rise to small  bumps and ripples in the surface when $\mathcal{K}^+$ is highly positive, or small dimples when $\mathcal{K}^+$ is highly negative.

\begin{figure}
\setlength{\unitlength}{0.003\columnwidth}
\begin{picture}(400,370)
\put(45,163){\includegraphics[width=0.65\columnwidth, keepaspectratio]{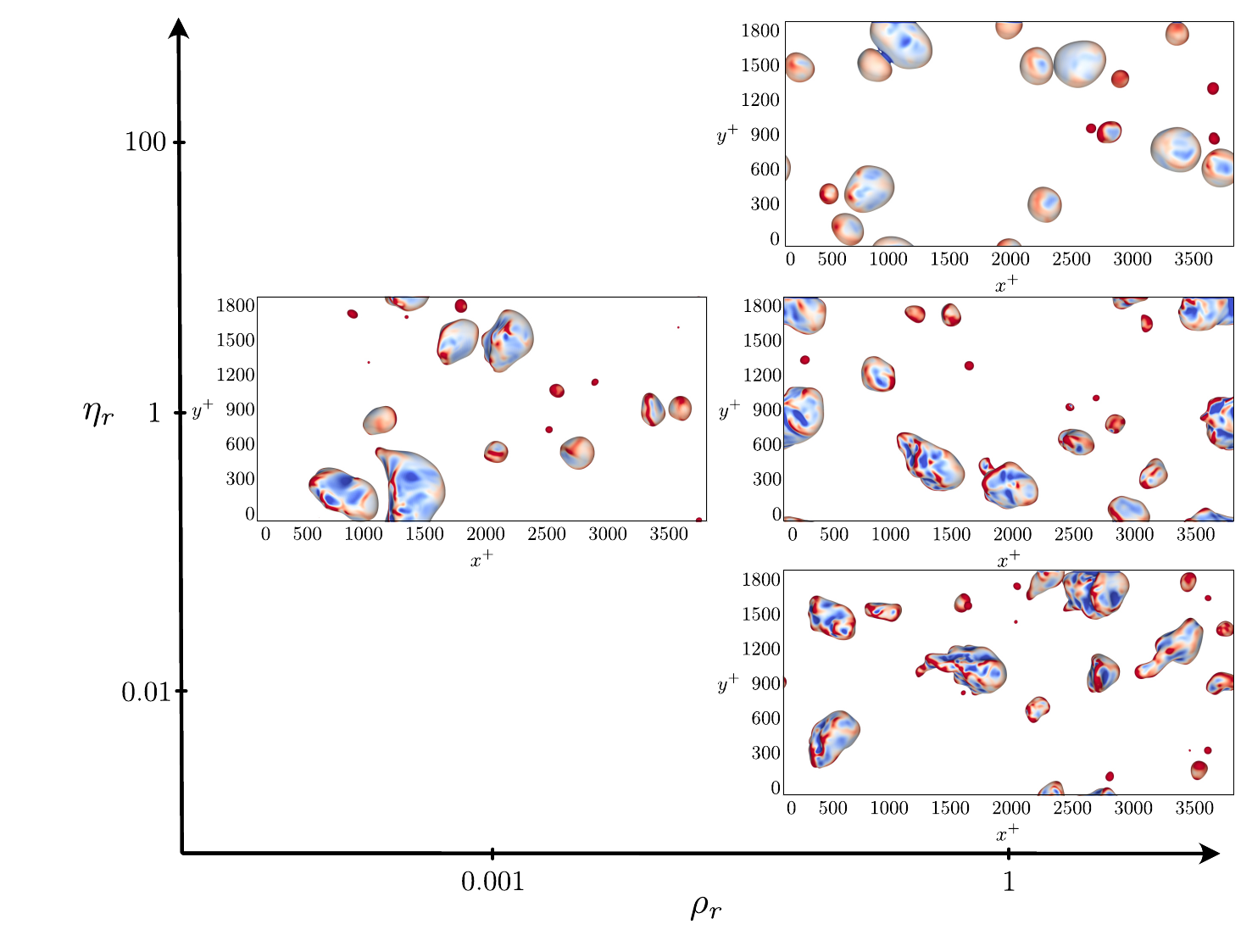}}
\put(45,-7){\includegraphics[width=0.65\columnwidth, keepaspectratio]{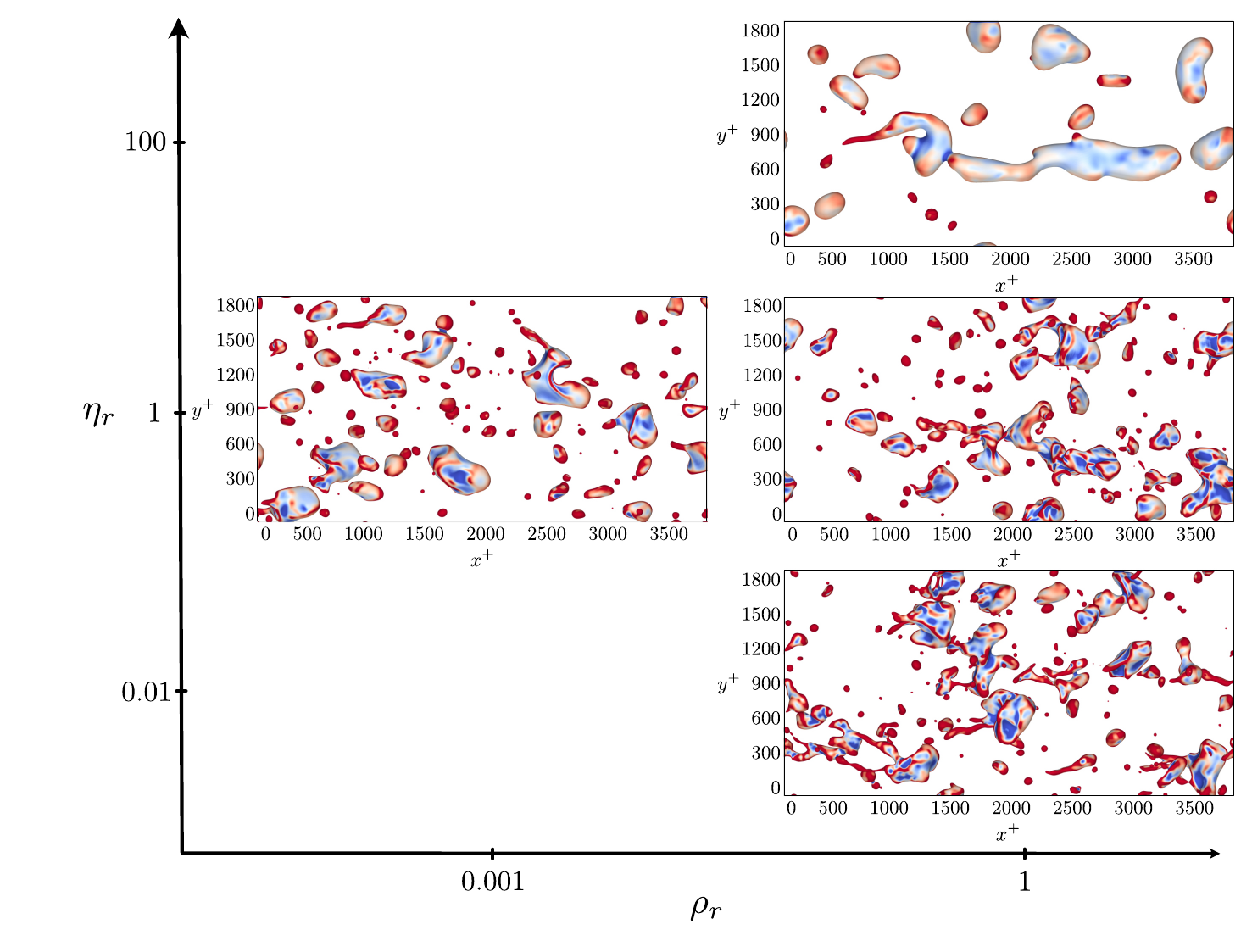}}
\put(230,61){\includegraphics[width=0.26\columnwidth, keepaspectratio]{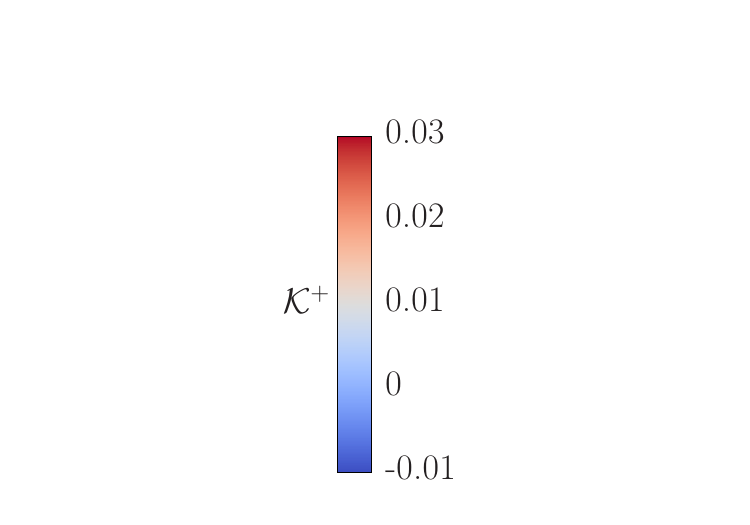}}
\put(230,231){\includegraphics[width=0.26\columnwidth, keepaspectratio]{5c}}
\put(60,315){(\textit{a})}
\put(60,145){(\textit{b})}
\put(110,300){$We=1.5$}
\put(110,130){$We=3.0$}
\end{picture}
\caption{Top view of the mean curvature of the bubble surface, $\mathcal{K}^+$, for four different combinations of density ratios ($\rho_r= 0.001$ and $1$) and viscosity ratios ($\eta_r= 0.01,1$ and $100$) once a statistically-stationary configuration is reached ($t^+=4000$).
Panel (\textit{a}) refers to $We=1.5$ while panel (\textit{b}) to $We=3.0$. 
The sub-panels are arranged in a plot using $\rho_r$ as $x$-coordinate and $\eta_r$ as $y$-coordinate. 
The effect of density can be appreciated in the sequence of panels on the middle row, while that of viscosity in the right column.
Bubble surface (iso-level $\phi=0$) is colored according to the local value of the mean curvature (low-blue; high-red).}
\label{fig:M}
\end{figure}

From figure~\ref{fig:M}, we can appreciate the effect of density and viscosity on the mean curvature from a qualitative point of view. 
The figure shows for $We=1.5$ (figure~\ref{fig:M}\textit{a}) and $We=3.0$ (figure~\ref{fig:M}\textit{b}) four top views of the statistically-stationary configurations of the system.
Bubbles are colored according to the local value of the mean curvature (blue-low; red-high).
Red areas correspond to bumps and ripples of the interface (positive curvatures), while blue areas to dimples (negative curvatures). 

For $We=1.5$  (figure~\ref{fig:M}\textit{a}), the effect of the density ratio can be observed by looking at the horizontal sequence of pictures (central row): we notice that moving from $\rho_r=1$ down to $\rho_r=0.001$ there is a slight decrease in the extension of both red and blue saturated regions, which correspond to very high and very low curvatures respectively.
Therefore a reduction of the density ratio (i.e. a decrease of bubble density), leads to a smoother bubble surface, characterized by fewer ripples and dimples.
In the vertical sequence of pictures on the right column, we can appreciate the effect of viscosity. 
We notice that the shape of the bubbles is qualitatively unchanged increasing the viscosity from $\eta_r=0.01$ to $\eta_r=1$.
However, from $\eta_r=1$ to $\eta_r=100$ the shape changes remarkably: the irregularities that characterize the bubbles surface at $\eta_r=1$ disappear completely at $\eta_r=100$, where the surface becomes very smooth and the bubbles shape very closely resembles the spherical shape. 
Thus, the action of viscosity seems opposite to the one of density in terms of local deformation of the bubble surface: an increase of viscosity prevents the formation of high curvatures values (in magnitude), while an increase of density promotes the formation of large interface deformations.
The two opposite trends obtained increasing the density or viscosity ratios can be interpreted in terms of local Reynolds or capillary numbers (i.e. evaluated using the bubble proprieties).
An increase of the density ratio leads to an increase of the local Reynolds number and as a consequence, a more irregular surface of the bubbles is obtained.
In contrast, an increase of the viscosity ratio, produces a decrease of the local Reynolds number (which also corresponds to an increase of the capillary number) and a smoother surface of the bubbles is attained.
Interestingly, the entity of these effects depends on the value of the ratio considered: a slight effect of the density ratio can be observed when it is decreased of three orders of magnitude (from $\rho_r=1$ down to $\rho_r=0.001$), as well as for the viscosity ratio when reduced by two orders of magnitude (from $\eta_r=1$ down to $\eta_r=0.01$), while a more noticeable difference is visible when it is increased of two orders of magnitude (from $\eta_r=1$ up to $\eta_r=100$).
Similar considerations can be obtained from the qualitative results obtained at $We=3.0$ (figure~\ref{fig:M}\textit{b}). 
In this case, we can qualitatively appreciate similar effects for the density and viscosity ratios.
These modifications, however, are now reflected on a much larger number of bubbles (larger Weber number).

\begin{figure}
\setlength{\unitlength}{0.003\columnwidth}
\begin{picture}(400,370)
\put(238,365){$We=3.0$}
\put(93,365){$We=1.5$}
\put(8,250){\includegraphics[width=0.51\columnwidth, keepaspectratio]{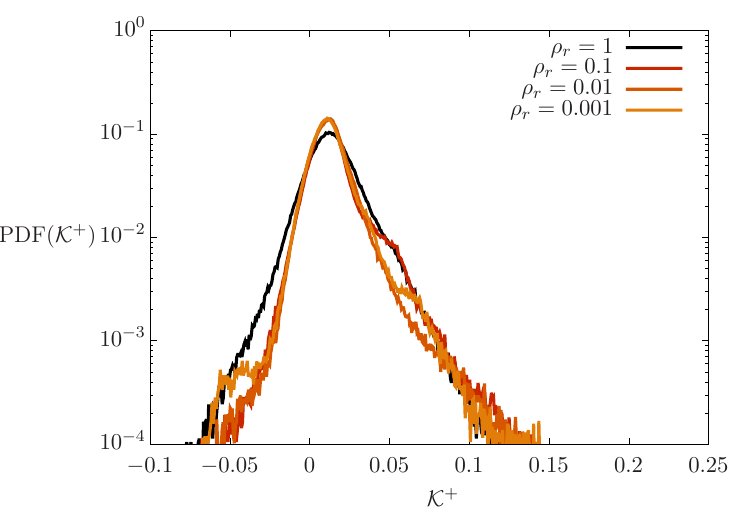}}
\put(154,250){\includegraphics[width=0.51\columnwidth, keepaspectratio]{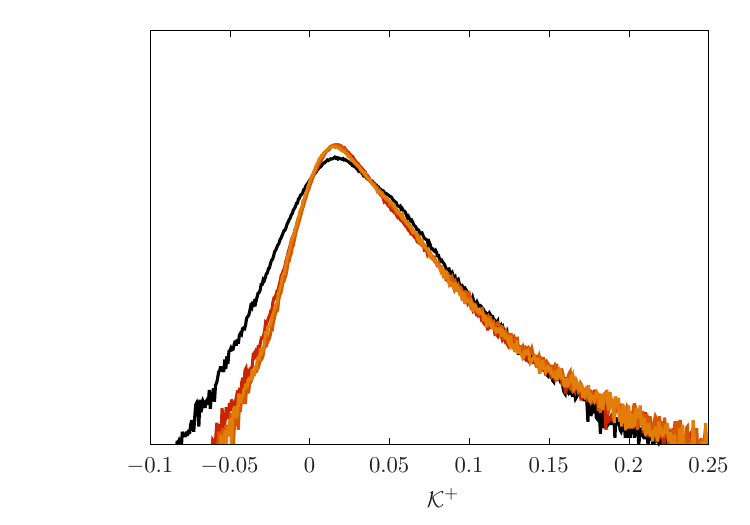}}
\put(124,299){\includegraphics[width=0.13\columnwidth, keepaspectratio]{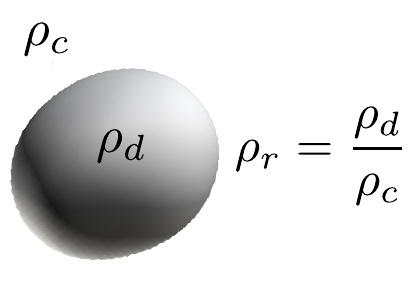}}
\put(24,355){(\textit{a})}
\put(178,355){(\textit{b})}
\put(238,240){$We=3.0$}
\put(93,240){$We=1.5$}
\put(8,125){\includegraphics[width=0.51\columnwidth, keepaspectratio]{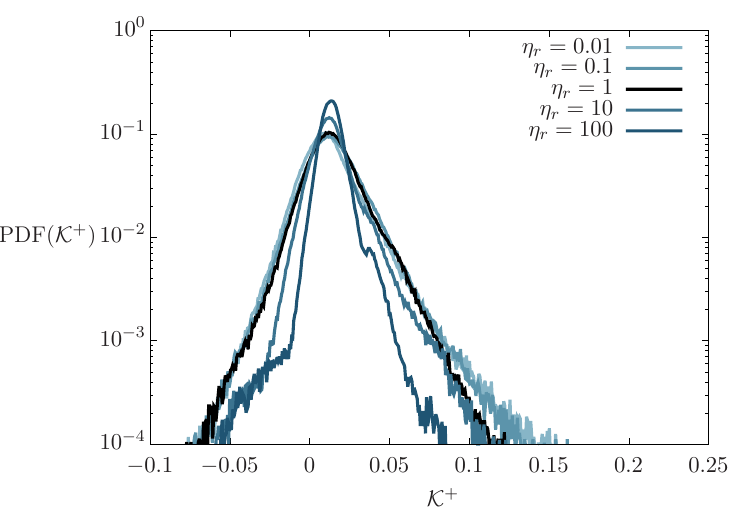}}
\put(154,125){\includegraphics[width=0.51\columnwidth, keepaspectratio]{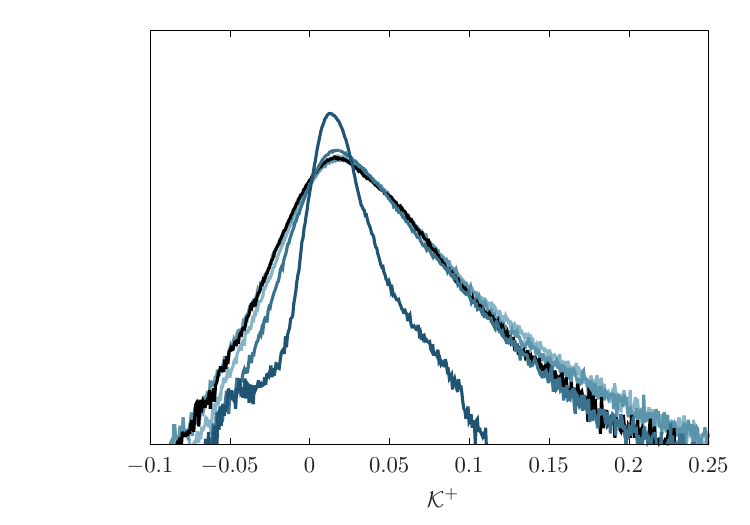}}
\put(124,167){\includegraphics[width=0.13\columnwidth, keepaspectratio]{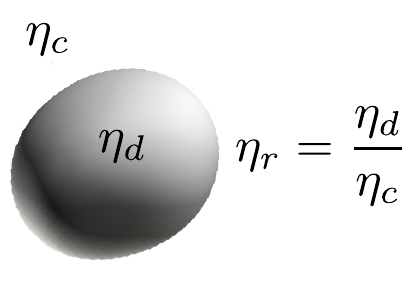}}
\put(24,230){(\textit{c})}
\put(178,230){(\textit{d})}
\put(238,115){$We=3.0$}
\put(93,115){$We=1.5$}
\put(8,0){\includegraphics[width=0.51\columnwidth, keepaspectratio]{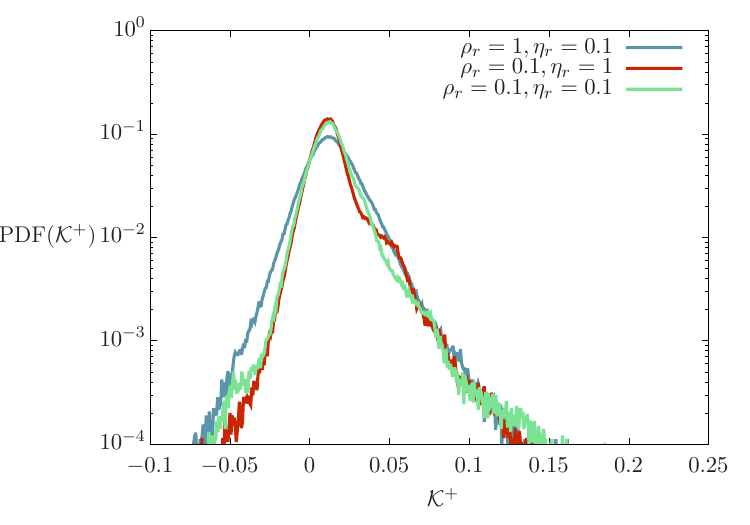}}
\put(154,0){\includegraphics[width=0.51\columnwidth, keepaspectratio]{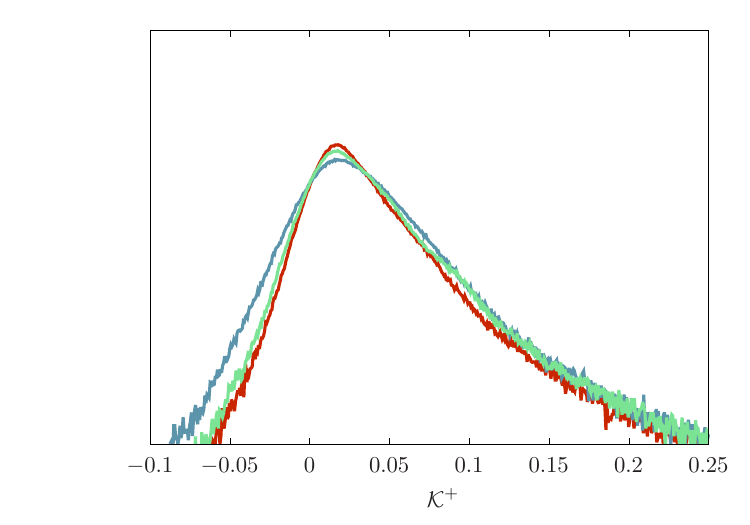}}
\put(124,50){\includegraphics[width=0.13\columnwidth, keepaspectratio]{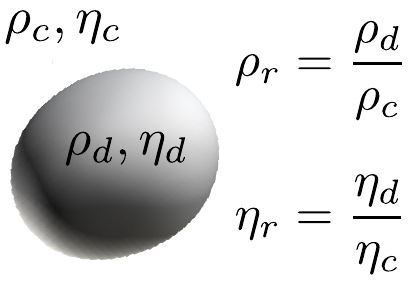}}
\put(24,105){(\textit{e})}
\put(178,105){(\textit{f})}
\end{picture}
\caption{Probability density function of the mean curvature, $\mathcal{K}^+$. 
Left column refers to $We=1.5$, while right column  to $We=3.0$. 
Effect of density ratio can be appreciated on the top row for $\rho_r=0.001, 0.01, 0.1$ and $1$ (with $\eta_r=1$).
The effect of bubble viscosity can be observed in the middle row for $\eta_r=0.01, 0.1, 1, 10$ and $100$ (with $\rho_r=1$).
Finally, the combined effect of the density and viscosity ratio is shown on the bottom row for the case with $\rho_r=0.1$, $\eta_r=0.1$, with respect to the cases where a single effect is considered (with $\rho_r=0.1$, $\eta_r=1$ and $\rho_r=1$, $\eta_r=0.1$).} 
\label{fig:Kpdf}
\end{figure}

To confirm these first qualitative observations, we compute the probability density function (PDF) of the mean curvature.
Results are reported in figure~\ref{fig:Kpdf} for different combinations of the density ratio, viscosity ratio, and Weber number.
Left column (figure~\ref{fig:Kpdf}\textit{a},\textit{c},\textit{e}) refers to $We=1.5$, while right column (figure~\ref{fig:Kpdf}\textit{b},\textit{d},\textit{f}) to $We=3.0$. 
Before analyzing each curve in detail, we can do some general observations. 
All curves are centered on a positive value of curvature and present an asymmetry with respect to the null value. 
Since positive curvatures correspond to convex surfaces and the null curvature corresponds to a flat surface, this is consistent with the fact that bubbles are in average convex, considering an outwards normal vector. 
Then, comparing the results shown in the left column (cases at $We=1.5$) against those reported in the right column (cases at $We=3.0$), we can appreciate the effect of the Weber number: for $We=3.0$ the curves are extended on a wider range of curvature values with respect to $We=1.5$. 
In particular, the curves are extended slightly towards negative values and considerably towards positive values, meaning that a higher Weber leads to a higher probability of having irregularities in the  surface of the bubbles, especially bump or ripples-like irregularities. 
The higher probability of having large curvature values is also due to the presence of many smaller bubbles, which are intrinsically more convex (smaller radius) and closer to a spherical shape.

We study now the effects of the density ratio (figure~\ref{fig:Kpdf}\textit{a},\textit{b}). 
We notice a trend for $We=1.5$ that becomes clearer for $We=3.0$: the cases with $\rho_r=0.1, 0.01, 0.001$ present a lower probability of having large curvatures (in magnitude) with respect to $\rho_r=1$. 
This effect is small for positive curvatures and more pronounced for negative curvatures. 
We can also observe that while the discrepancy between the reference case ($\rho_r=1$) and all other cases is clear, there is almost no difference among the cases $\rho_r=0.1, 0.01,$ and $0.001$.   
Interestingly, a similar trend was also reported in a previous work \cite{cano2015use} that investigated the rise of bubbles in quiescent liquid.
In particular, \citet{cano2015use} reported that for density ratios smaller than 0.128, a further decrease of the density ratio does not produce significant changes in the shape of the bubbles.
This seems to suggest that the modifications produced by the density with respect to the case with $\rho_r=1$ (matched density case), are likely to be proportional to the density difference between the two phases (i.e. $\rho_c -\rho_d$) rather than their ratio (i.e. $\rho_d/\rho_c$).
Further simulations, which consider super-unitary density ratios, are however required to confirm this indication.
Overall, present results (figure~\ref{fig:AI}) indicate that when sub-unitary density ratios are considered, the probability of having large curvatures values, especially negative, and very stretched bubbles decreases. 
In other words, when the density of the bubbles is decreased with respect to the carrier density, it becomes more difficult for turbulence fluctuations to locally deform and stretch the bubbles, and in particular, it is difficult to create dimples and concave areas. 
A possible physical mechanism that supports present observations is the following: when an external perturbation reaches the deformable interface of a bubble, the bubble surface is modified and the perturbation then propagates to the internal bubble fluid.
As bubble density is reduced, however, the propagation of this perturbation to the bubble fluid and thus to the rest of the bubble interface becomes less effective.
Indeed, the inertia of the perturbation is modulated by the smaller bubble density and thus the magnitude of the inertial forces is reduced.
As a result, viscous and surface tension forces increase their relative importance with respect to inertial forces, and the resulting bubble deformation is reduced.
This behavior can be also justified considering the dispersed phase Reynolds number, i.e. the Reynolds number evaluated considering the dispersed phase density.
As bubble density is reduced, so does the dispersed phase Reynolds number and the bubbles become less deformable and distorted, as can be also graphically appreciated from figure~\ref{fig:T} comparing the case $\rho_r=0.001$ (orange bubbles) against the case $\rho_r=1.000$ (white bubbles).



To evaluate the influence of the viscosity, we consider figure~\ref{fig:Kpdf}\textit{c},\textit{d}.
A trend can be distinguished for both the Weber numbers: the PDFs become narrower as the viscosity increases. 
More specifically, the largest effect can be seen for $\eta_r=100$, where the range of possible curvatures is significantly reduced. 
The shrinkage of the pdf is less but still evident for $\eta_r=10$, and it becomes almost negligible for $\eta_r=0.1$ and $\eta_r=0.01$.
Unlike density, the impact of viscosity is important for $\eta_r=100$ and $\eta_r=10$, while it becomes less important for $\eta_r=0.1$ and $\eta_r=0.01$.
Indeed, for these two latter cases, no significant modifications can be appreciated from both Weber numbers.

Finally, the combined effects of the density and viscosity ratio can be evaluated from figure~\ref{fig:Kpdf}\textit{e},\textit{f}.
Interestingly, we observe that when both density ratio and viscosity ratios are decreased, the resulting PDF of the mean curvature lies in between the case $\rho_r=0.1$ (and matched viscosity) and $\eta_r=0.1$ (and matched density).
This intermediate behavior can be traced back to the two opposite actions of density and viscosity on the mean curvature of the surface of the bubbles: while a decrease of the bubble density (i.e. of the density ratio) makes the bubbles surface more rigid and thus smoother, when bubble viscosity is decreased the bubbles become more deformable and ripples or dimples can be more easily formed on the interface.
Thus, when we combine these two effects, these actions balance out and we obtain an intermediate trend.
This result is already visible for $We=1.5$ and becomes clearer for $We=3.0$ where, thanks to the higher number of bubbles, a smoother statistic is obtained.

\subsection{Flow modifications}
 
\subsubsection{Mean velocity profiles}

Once detailed the evolution of the dispersed phase topology, its modifications and the deformation and curvature of the bubbles, we move to analyze the flow modifications produced by the bubbles.
We start by analyzing the macroscopic behavior of the multiphase mixture, in terms of flow-rate and mean flow statistics. 
In particular, we investigate the wall-normal behavior of the mean velocity profiles of the multiphase flow, and we compare them with the single-phase flow statistics at the same $Re_\tau=300$.
Indeed, we aim at understanding whether the injection of bubbles in a turbulent flow induces modifications to the mean velocity profile, especially when density or viscosity contrasts are present between the two phases. 
This aspect is widely studied and a common question that persists in the field concerns the capability of bubbles in generating drag reduction \cite{LU2005,CECCIO2010,VANGILS2013,verschoof2016bubble,Scarbolo2016,cannon2021effect}.

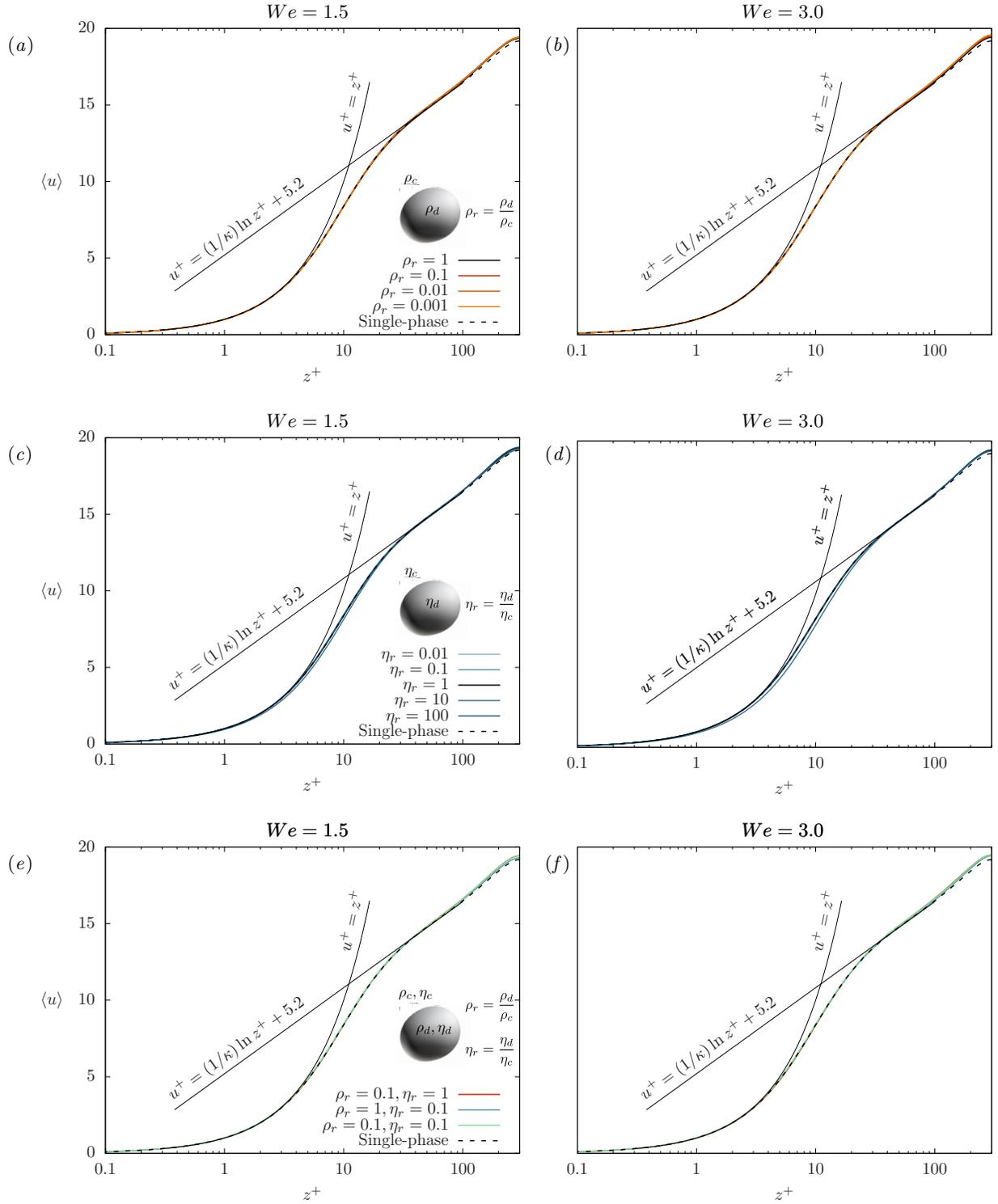
\begin{figure}
\setlength{\unitlength}{0.003\columnwidth}
\begin{picture}(400,370)
\put(238,365){$We=3.0$}
\put(93,365){$We=1.5$}
\put(10,250){\includegraphics[width=0.51\columnwidth, keepaspectratio]{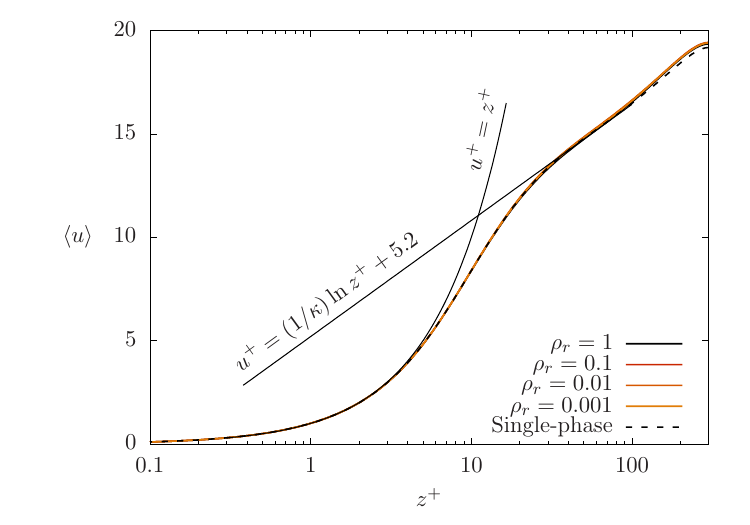}}
\put(154,250){\includegraphics[width=0.51\columnwidth, keepaspectratio]{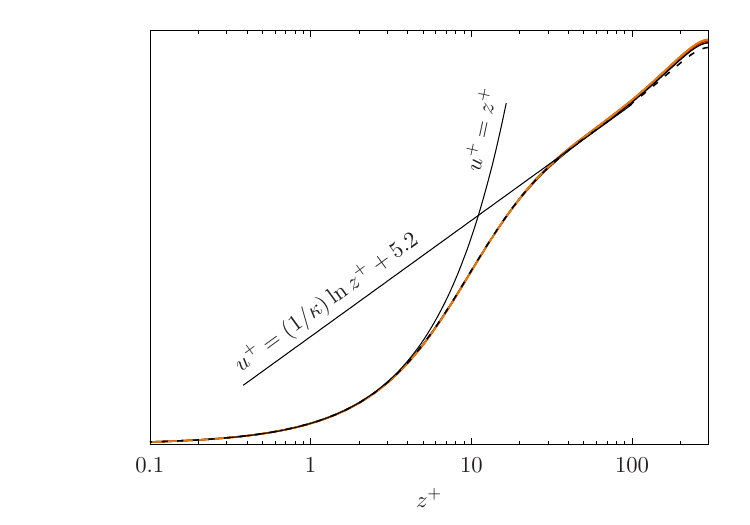}}
\put(133,293){\includegraphics[width=0.11\columnwidth, keepaspectratio]{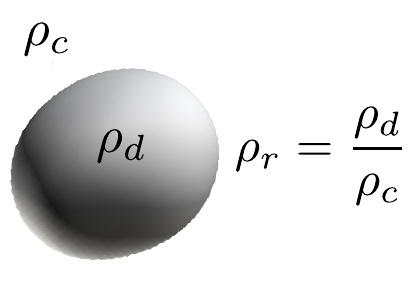}}
\put(14,355){(\textit{a})}
\put(178,355){(\textit{b})}
\put(238,240){$We=3.0$}
\put(93,240){$We=1.5$}
\put(10,125){\includegraphics[width=0.51\columnwidth, keepaspectratio]{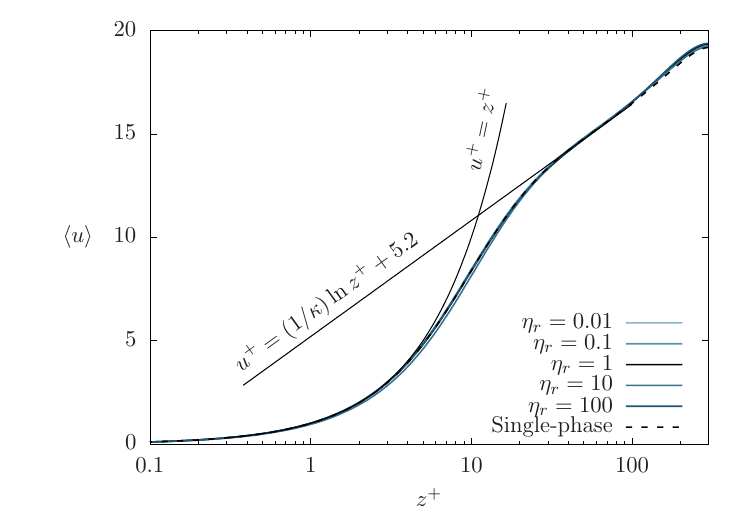}}
\put(154,124){\includegraphics[width=0.51\columnwidth, keepaspectratio]{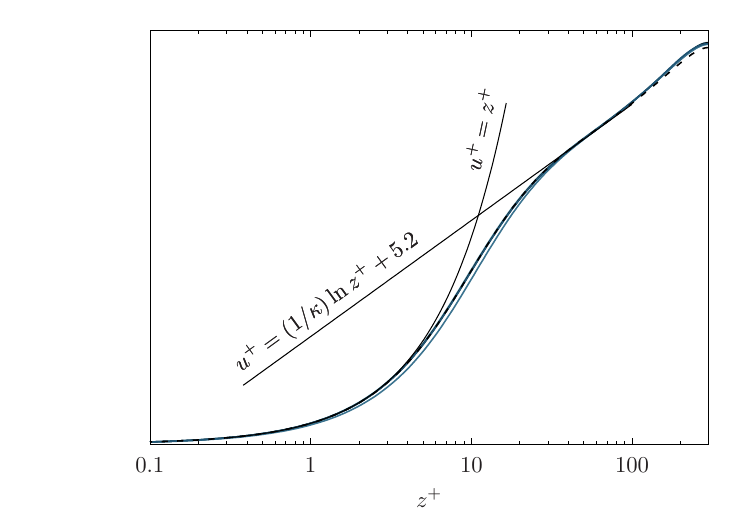}}
\put(133,173){\includegraphics[width=0.11\columnwidth, keepaspectratio]{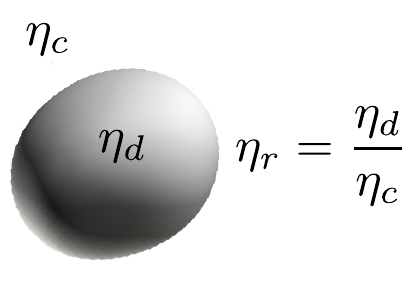}}
\put(14,230){(\textit{c})}
\put(178,230){(\textit{d})}
\put(238,115){$We=3.0$}
\put(93,115){$We=1.5$}
\put(10,0){\includegraphics[width=0.51\columnwidth, keepaspectratio]{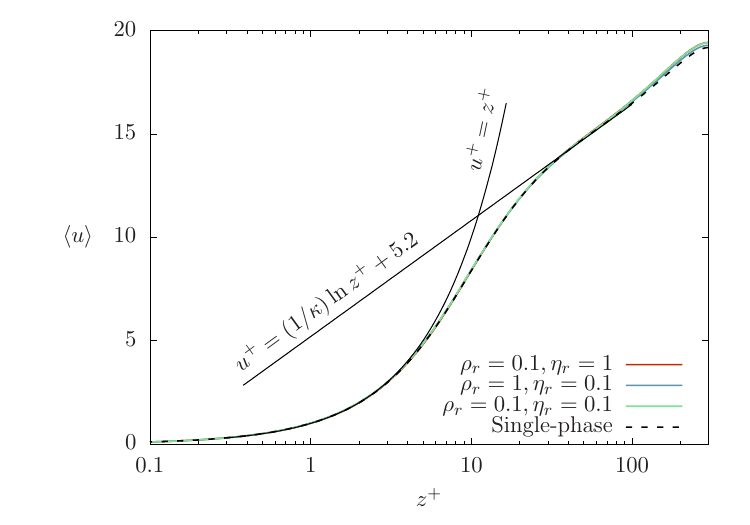}}
\put(154,0){\includegraphics[width=0.51\columnwidth, keepaspectratio]{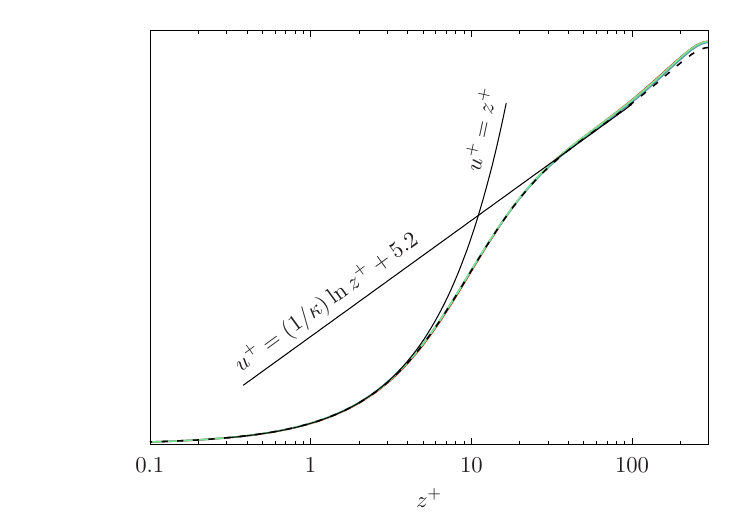}}
\put(133,43){\includegraphics[width=0.11\columnwidth, keepaspectratio]{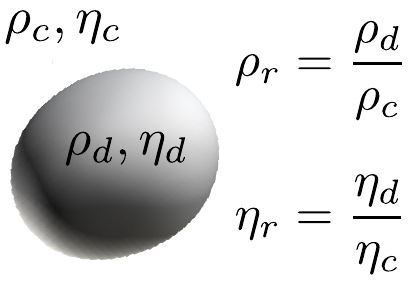}}
\put(14,105){(\textit{e})}
\put(178,105){(\textit{f})}
\put(238,115){$We=3.0$}
\put(93,115){$We=1.5$}
\end{picture}
\caption{Wall-normal behavior of the streamwise mean velocity profiles.
Left column refers to $We=1.5$, while the right column  to $We=3.0$. 
Density ratios effects are shown on the top row for $\rho_r=0.001, 0.01, 0.1, 1$.
Viscosity ratio effects are shown on the middle row for $\eta_r=0.01, 0.1, 1, 10, 100$.
Finally, the combined effect of the density and viscosity ratios is shown on the bottom row for the case $\rho_r=0.1$ and $\eta_r=0.1$, with respect to the cases where only one effect is considered.
As a reference, the classical law of the wall, $u^+ = z^+$ and $u^+ = (1/k ) \log z^+ + 5$ (with $k = 0.41$ the von K\'{a}rm\'{a}n constant) is also reported with a dashed line.
For all cases, with respect to single-phase, we observe a minor increase of the mean velocity.} 
\label{fig:Umean}
\end{figure}

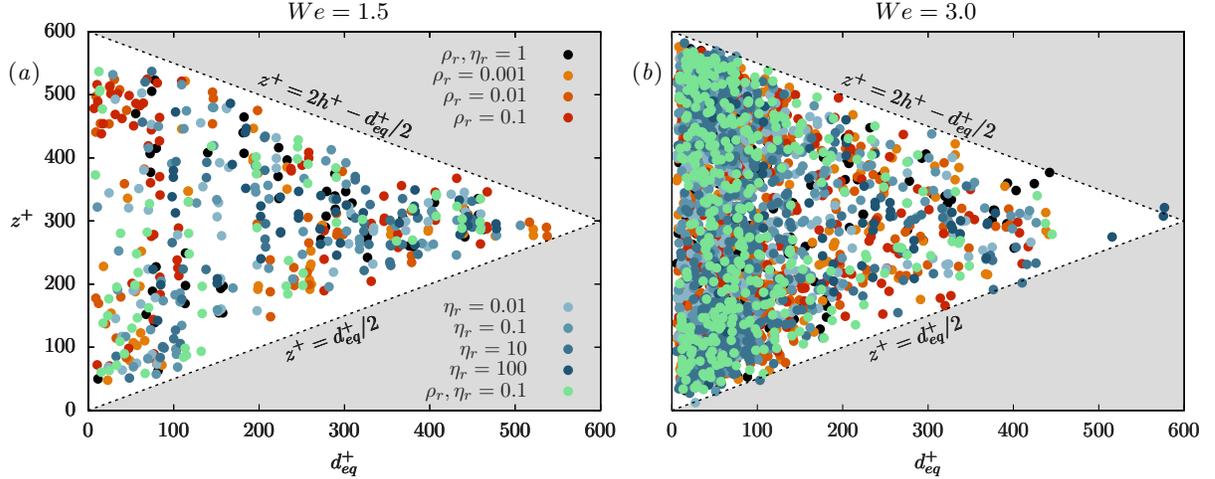
\begin{figure}
\setlength{\unitlength}{0.003\columnwidth}
\begin{picture}(400,130)
\put(238,115){$We=3.0$}
\put(93,115){$We=1.5$}
\put(10, 0){\includegraphics[width=0.51\columnwidth,keepaspectratio]{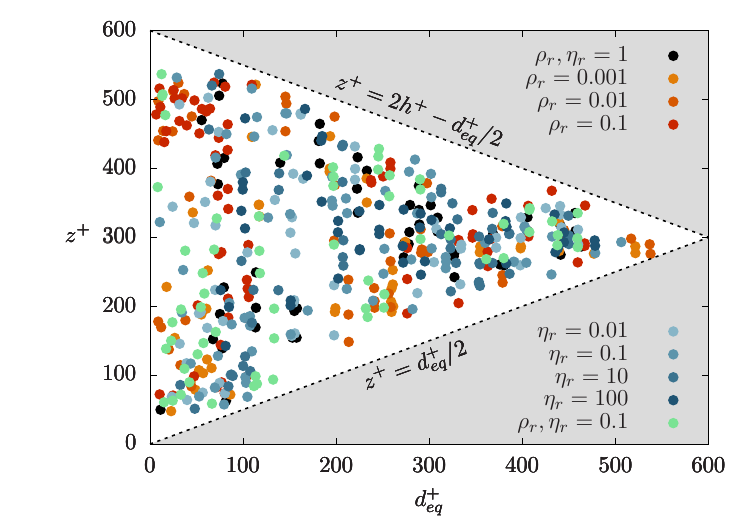}}
\put(154,0){\includegraphics[width=0.51\columnwidth, keepaspectratio]{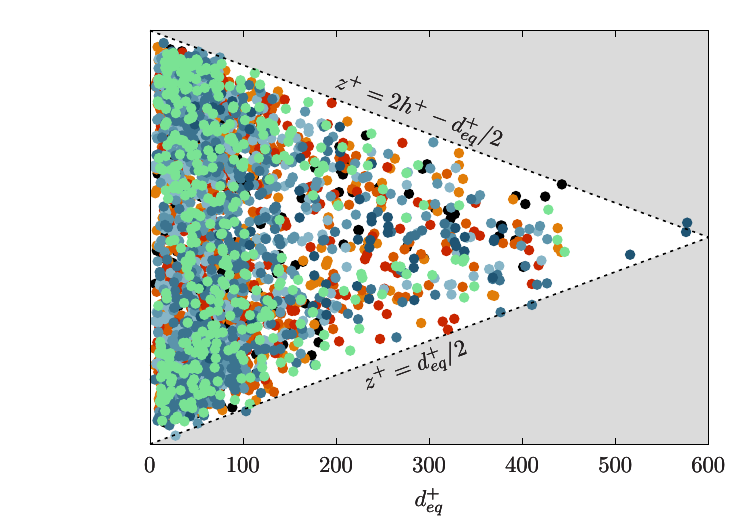}}
\put(24,100){(\textit{a})}
\put(178,100){(\textit{b})}
\end{picture}
\caption{Scatter plot of the wall-normal location of each bubble over its size for the different cases considered.
The two black dashed lines identify the condition for which the interface of the bubble intercepts the closer wall in the hypothesis of a perfectly spherical bubble.
Smaller bubbles tend to disperse along the entire channel height can get rather close to one of the two walls while larger bubbles tend to accumulate at the center of the channel.}
\label{scatter}
\end{figure}

Figure~\ref{fig:Umean} shows the wall-normal behavior of the mean velocity profiles, computed by averaging the streamwise velocity along the streamwise and spanwise directions in the entire domain (both dispersed and carrier phase). 
The results are illustrated for all combinations of density and viscosity ratios considered, following the same arrangement of the previously presented statistics. 
In addition, the velocity profile relative to the single-phase case is shown with a black dashed line, and the classical law of the wall, $u^+=z^+$ and $u^+=(1/\kappa) \ln z^+ + 5.2$ \cite{POPE2000}, is reported as a reference (with $\kappa=0.41$ the Von Kármán constant \cite{von1931mechanical}). 
We observe that in all the plots the velocity profiles perfectly collapse on each other in the vicinity of the wall, while tiny deviations can be observed in the central part of the channel, where most bubbles are located. 
In particular, in the core region of the channel, no differences can be appreciated varying the density and viscosity ratios.
However, all multiphase cases are characterized by a slightly greater velocity with respect to the single-phase case. 
As in our simulations a constant mean pressure gradient is used to drive the flow, the observed flow-rate enhancement corresponds to a slight drag reduction.
The drag reduction we observe is rather low in all the simulated cases (roughly 1 to 2\%), and current results suggest that the presence of density and viscosity contrasts among the phases does not visibly impact it. 
These results are in agreement with previous works \cite{Lu2019,cannon2021effect}, which found that drag significantly depends on the bubble size.
Specifically, they observe that large and deformable bubbles (obtained allowing bubbles to coalesce) migrate towards the central part of the channel and do not influence the drag significantly \cite{LU2005,Lu2007,Lu2008,LuMT_2017}.
By opposite, smaller bubbles (obtained not allowing bubbles to coalesce) move towards the near-wall region and lead to an increase of the drag \cite{LU2005,Lu2007,Lu2008,LuMT_2017}. 
To support this argument, we can consider figure~\ref{scatter}, which shows the scatter plot of the wall-normal location of each bubble over its equivalent diameter.
Panel~\textit{a} refers to $We=1.5$ while panel~\textit{b} to $We=3.0$.
The bottom and top walls are located at $z^+=0~w.u.$ and $z^+=600~w.u.$.
Two black dashed lines identify the critical condition for which the upper (or lower) part of the bubble interface intercepts the top (or bottom) wall.
From a mathematical point of view, this condition can be identified imposing:
\begin{equation}
z_b^+ = d_{eq}^+/2\, ,
\end{equation}
where $z_b^+$ is the distance of the center of the mass of the bubble from the closer wall, which can be computed as follows:
\begin{equation}\
z_b^+=\text{min}(z_i^+, 2h^+ -z_i^+)\, ,
\end{equation}
where $z_i^+$ is the wall-normal location of the $i$-th bubble and $h^+=300~w.u.$ is the channel half-height in wall units.
Hence, the equations that identify these conditions are: 
\begin{equation}
z^+=d_{eq}^+/2\, ,  \qquad z^+=2h^+ -d_{eq}^+/2\, .
\end{equation}
Analyzing the dispersion of the bubbles along the wall-normal direction, we can confirm previous intuitions: smaller bubbles tend to disperse along the entire height of the channel and can get rather close to the two walls while, by opposite, larger bubbles tend to accumulate at the center of the channel and stay farther away from the two walls.
It is worth pointing that despite a few points are located above (or below) the two black dashed lines (i.e. in the gray region), no collisions with the walls are detected. 
Instead, these points represent bubbles elongated along the streamwise or spanwise directions and thus with a larger $d_{eq}^+$ with respect to the actual wall-normal size.
Overall, the results presented in figure~\ref{fig:Umean} corroborated by those reported in figure~\ref{scatter} seem to confirm the idea that bubble deformability is a crucial parameter for obtaining drag reduction \cite{Segre1962a,Lu2007,CECCIO2010,verschoof2016bubble,cannon2021effect}.
Indeed, bubble deformability plays a central role in determining the preferential distribution of the bubbles \cite{Lu2007,Lu2018}, which is directly linked to drag reduction \cite{verschoof2016bubble,Scarbolo2016}.

\subsubsection{Turbulent Kinetic Energy (TKE) of bubbles}

\begin{figure}
\begin{picture}(420,350)
\put(10,0){\includegraphics[width=0.78\columnwidth, keepaspectratio]{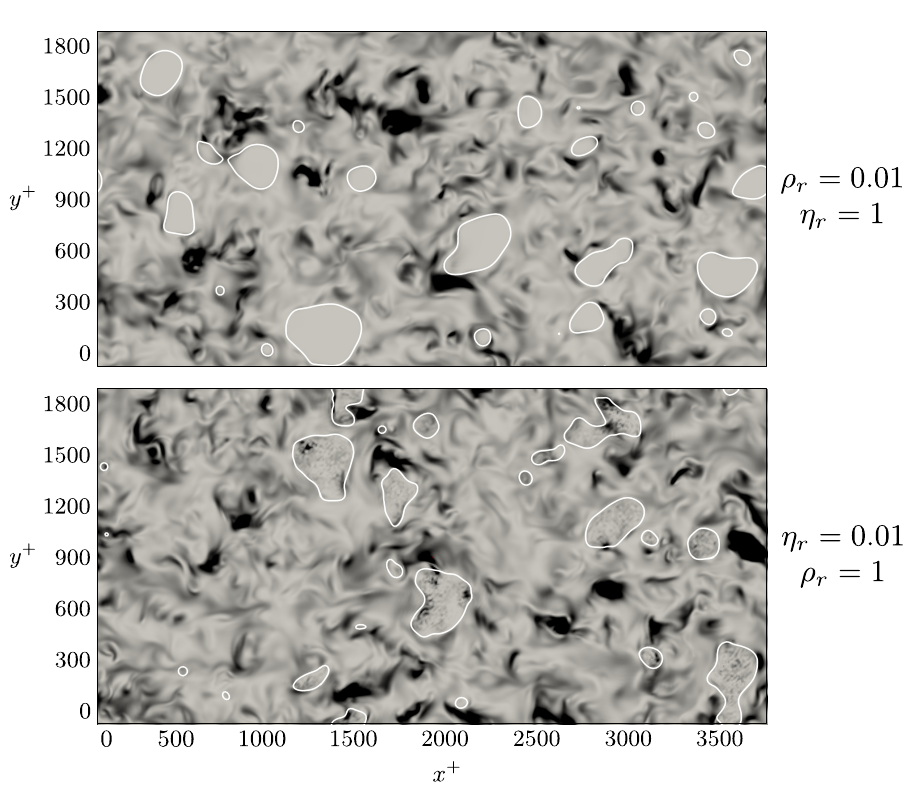}}
\put(285,125){\includegraphics[width=0.38\columnwidth, keepaspectratio]{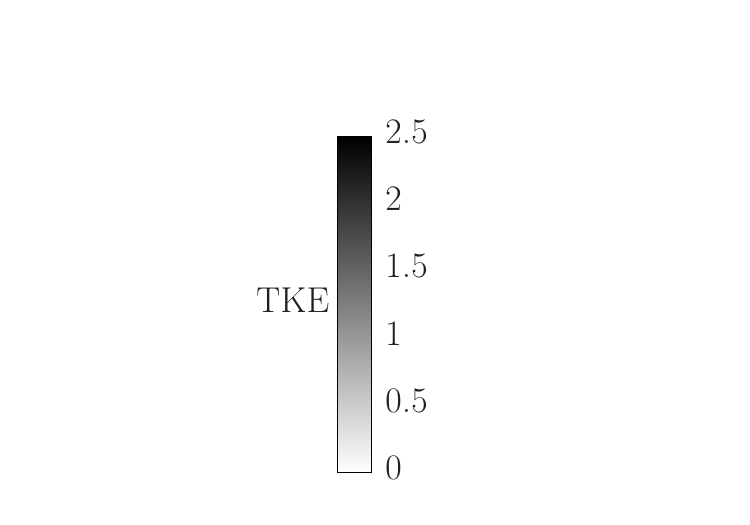}}
\put(1,325){(\textit{a})}
\put(1,170){(\textit{b})}
\end{picture}
\caption{Contour map of the turbulent kinetic energy in a $x^+-y^+$ plane located at the channel center ($z^+=0$).
Panel (\textit{a}) refers to the case $\rho_r=0.01$ and $\eta_r=1$ while panel (\textit{b}) refers to the case $\rho_r=1$ and $\eta_r=0.01$.
Both panels refer to the lower surface tension case ($We=3.0$) and to the time instant $t^+ = 4000$ (statistically-steady configuration). 
The interface of the bubbles is highlighted with a white line.
For $\rho_r=0.01$, bubbles and characterized by a low and uniform value of the TKE while, for $\eta_r=0.01$, the TKE map is  non-uniform and characterized by small scales fluctuations.}
\label{fig:tkeimg}
\end{figure}

After having analyzed the flow field in terms of mean velocity, we focus on the turbulence behavior inside the bubbles. 
The characterization of the flow inside the bubbles is of paramount importance in many applications.
Indeed, internal circulation controls the transport of heat, mass, momentum and chemical species through the interface \cite{ayyaswamy1990effect,Gissinger2021}, the motion and deformation of the bubbles \cite{levich1963,mashayek1998nonlinear} and particle removal efficiency in scrubbing process \cite{Ghiaasiaan1997,Hajisharifi2021}.
To characterize the mixing and flow behavior in the dispersed phase, we consider the turbulent kinetic energy (TKE) inside the bubbles.
As in the carrier phase no significant modifications of the mean velocity profile (figure~\ref{fig:Umean}) and of turbulence statistics are observed, larger modifications are expected in the dispersed phase: the flow inside the bubbles is confined by a deformable interface and continuously forced by the external carrier flow.
In addition, fluid properties (density and viscosity) are different.
As a results, the magnitude of inertial and viscous forces is changed, as well as the local Reynolds and Weber numbers (i.e. evaluated using the dispersed phase properties).
To give a first qualitative idea of these modifications, we can consider the specific turbulent kinetic energy, TKE, whose definition is here recalled:
\begin{equation}
\text{TKE}=\frac{\rho}{\rho_c}\frac{(u'^2+v'^2+w'^2)}{2}\, , 
\label{stke}
 \end{equation}
 where $\rho$ is the local density ($\rho_d$ in the bubbles and $\rho_c$ in the carrier phase).
Figure~\ref{fig:tkeimg} shows the turbulent kinetic energy for two different simulations: panel ({\textit{a}}) refers to the case with $\rho_r=0.01$ and matched viscosity and panel ({\textit{b}}) to the case with $\eta_r=0.01$ and matched density.
Both panels refer to the higher Weber number analyzed ($We=3.0$) and to the time instant $t^+=4000$, when for both cases a statistically-stationary configuration is attained.
The two snapshots illustrate with a white-black scale the contour map of TKE on an $x^+-y^+$ plane located at the channel center ($z^+=0$).
The interface of the bubbles is marked with a white thin line. 
We notice that the flow structures in the carrier phase are qualitatively similar in the two pictures, while inside the bubbles the contour maps of TKE look very different and for $\rho_r=0.01$ and $\eta_r=1$ (panel {\textit{a}}), low values of TKE inside the bubbles. 
In evaluating the results presented in panel {\textit{a}}, however, it is important to make an important observation: although the energy content of the bubbles is rather low, velocity fluctuations are still present inside the bubbles.
Indeed, the low values of TKE in the bubbles obtained for the case $\rho_r=0.01$ and $\eta_r=1$ are due to the low density that characterizes the bubbles: the prefactor $\rho/\rho_c$ present in the definition of TKE reduces the values obtained inside the bubbles.
Shifting our focus to the case $\eta_r=0.01$ and $\rho_r=1$ (panel {\textit{b}}), we can appreciate here the presence of many vortical structures characterized by an energy content similar to that of the carrier phase.
Interestingly, the characteristic length scale of these turbulence structures is much smaller than that of the carrier phase.
This observation can be traced back to the smaller viscosity of the dispersed phase that results in a larger local Reynolds number, as also observed in other multiphase flow instances \cite{Roccon2019,roccon2021}.

\begin{figure}
\setlength{\unitlength}{0.003\columnwidth}
\begin{picture}(430,200)
\put(0,0){\includegraphics[width=0.55\columnwidth, keepaspectratio]{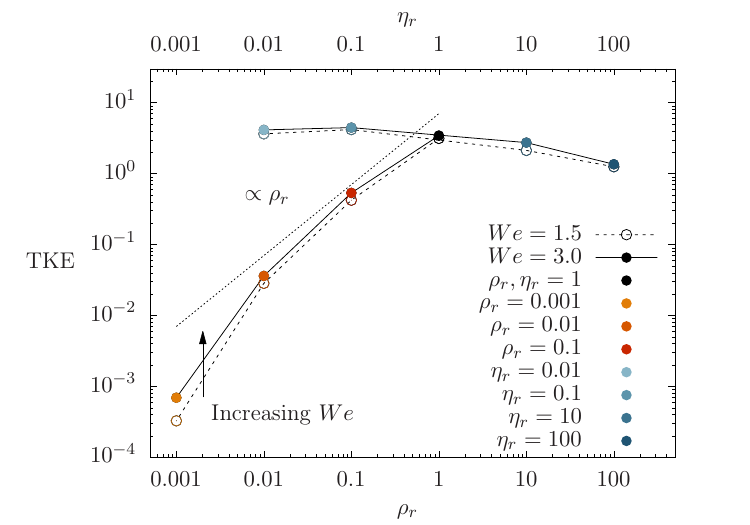}}
\put(164,0){\includegraphics[width=0.55\columnwidth, keepaspectratio]{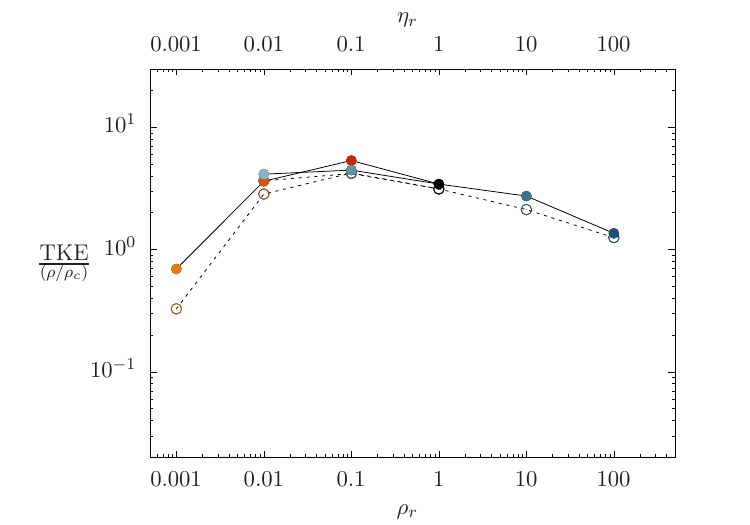}}
\put(14,100){(\textit{a})}
\put(178,100){(\textit{b})}
\end{picture}
\caption{Mean value of the turbulent kinetic energy (TKE) inside the bubbles.
In panel~\textit{a}, TKE is evaluated using the complete definition of specific TKE (i.e. including the prefactor $\rho/\rho_c$) while, in panel \textit{b}, TKE is evaluated considering only the velocity contribution (i.e. not considering  the prefactor $\rho/\rho_c$).
For both panels, a dashed line ($We=1.5$) and a continuous line ($We=3.0$) are used to show the behavior of TKE as the density or viscosity ratios are changed. 
Each value of TKE is marked with a circle (empty for $We=1.5$ and filled for $We=3.0$), with a red-color scale  for the non-matched density cases and a blue-color scale for the non-matched viscosity cases, while the black color is used for the reference case.}
\label{fig:tkeplot}
\end{figure}

Turbulence inside the bubbles is the mechanism that can increase or decrease transfer rates across the interface \cite{saboni2007effect,cano2015use}.
To quantify more closely this aspect, we compute the mean value of the specific turbulent kinetic energy inside the bubbles for all simulated cases, except for the combined case, and we collect the results  in figure~\ref{fig:tkeplot}. 
To better evaluate the contribution of density and velocity fluctuations in the resulting TKE values, turbulent kinetic energy is evaluated using the complete definition (equation~\ref{stke}) in panel~\textit{a} while TKE is evaluated considering only the velocity fluctuations contribution in panel~\textit{b}, (i.e. TKE is reported normalized by the local density contribution $\rho/\rho_c$).
The mean values of TKE are reported as a function of the density ratio (scale on the bottom part of the plot), viscosity ratio (scale on the top part of the plot), and Weber number (full circles for $We=1.5$ and empty circles for $We=3.0$).
We start by analyzing the effects of the density and viscosity ratios shown in panel~\textit{a}, we can observe two opposite trends: as the viscosity ratio increases, the mean value of TKE inside the bubbles decreases of about one order of magnitude while, by opposite, increasing the density ratio, the mean value of TKE inside the bubbles rapidly increases (of about four orders of magnitude).
This behavior reflects the modifications of the inertial and viscous forces inside the bubbles produced by the different dispersed phase density and viscosity.
As the viscosity ratio is increased from $\eta_r=0.01$ up to $\eta_r=100$ (from left to right), viscous forces become dominant over inertial forces and thus local Reynolds number decreases.
As a result, for low viscosity ratios, we observe small turbulent structures inside the bubbles characterized by significative TKE levels, while, for viscosity ratios larger than unity, turbulence structures cannot be sustained inside the bubbles (larger viscous dissipation) and bubbles are characterized by a low level of TKE.
A similar trend, albeit in a slightly different simulation setup, was reported by \citet{cano2015use} that investigated the rise of bubbles in still liquid and observed a reduction of the velocity gradients for increasing values of the viscosity ratio.
On the other hand, increasing the density ratio from $\rho_r=0.001$ up to $\rho_r=1$, inertial forces become dominant over viscous forces, the local Reynolds number increases and the bubbles are characterized by larger TKE values.
Interestingly, we observe a much stronger action of the density ratio on the mean value of the bubbles TKE.
Indeed, if we compute the specific turbulent kinetic energy using equation~\ref{stke}, the resulting TKE values directly depend on the bubble density and, as we can see from panel~\textit{a}, present results roughly follow the $\rho_r$ scaling law reported with a dotted line.
However, it is worthwhile observing that when the smallest density ratio is considered ($\rho_r=0.001$), results start to deviate from the $\rho_r$ scaling law: as the density ratio is reduced, we observe a reduction int the magnitude of the velocity fluctuations of about one order of magnitude.
This deviation can be better appreciated in panel~\textit{b}, where TKE values are reported normalized by the prefactor  $\rho/\rho_c$, so that the contribution from velocity fluctuations alone can be better appreciated. The magnitude of velocity fluctuations is roughly constant when considering different density ratios, exception made for the lowest density ration, $\rho_r=0.001$, thus indicating that the specific TKE scales with the density ratio.
Finally, we can consider the effect of the Weber number: increasing the Weber number, thus decreasing the surface tension, the TKE is slightly increased for all the cases. 
This trend can be attributed to the larger transfer of momentum that occurs when surface tension forces are weaker: as the interface becomes more deformable, the modulation effect of the interface becomes weaker and energy and momentum can be more easily exchanged between the phases.
When the surface tension is reduced, in fact, the bubbles become more deformable and reasonably they are more likely to contain a greater amount of TKE.


\section{Conclusions}
\label{sec: conclusions}

In this work, we studied the behavior of bubbles in a turbulent channel flow for different values of the density ratio, viscosity ratio, and Weber number.
The investigation is based on direct numerical simulation of turbulence coupled with a phase-field method.
First, we investigated the topology of the dispersed phase and its modifications. 
We found that the number of bubbles present in the channel is strongly influenced by the surface tension value (i.e. by the Weber number), in accordance with the results of previous studies \cite{Scarbolo2015,Roccon2017,Soligo2019c}.
Besides, we observe that an increase of bubble viscosity with respect to the carrier (i.e. an increase of the viscosity ratio) has an important stabilizing role and leads to a remarkable increase of the maximum bubble stable diameter and thus to a decrease of the number of bubbles.
By opposite, a reduction of the bubble density (i.e. a reduction of the density ratio), does not remarkably affect the dispersed phase topology.
Similar findings are obtained from the analysis of the coalescence and breakage rates: an increase of bubble viscosity or surface tension (i.e. a decrease of the Weber number) leads to a reduction of the breakage and coalescence rates.
In contrast, a modification of the density ratio has a marginal effect on the behavior of the breakage and coalescence rates.
Secondly, we studied the surface stretching and curvature of the bubbles.
We observed that these indicators are influenced by all three parameters investigated.
In particular, larger viscosity ratios or lower density ratios or Weber numbers hinder the stretching of the bubbles and as a result the overall amount of interfacial area obtained is lower (with respect to the reference matched density and viscosity cases).
These observations are also reflected in the probability density function of the mean curvature: an increase of bubble viscosity, a decrease of bubble density or a decrease of the Weber number hinder the formation of ripples and dimples on the surface of the bubbles and thus high curvature values are less likely to be found.
Thirdly, we evaluated the flow modifications produced by the swarm of bubbles in the background turbulent flow and in the dispersed phase.
From a macroscopic point of view, no significant modifications are observed in the wall-normal behavior of the mean velocity profiles and only a minor increase of the flow-rate is observed for all bubbles-laden cases with respect to a single-phase flow, in accordance with previous results \cite{Scarbolo2015,Roccon2017,Soligo2019c}.
Finally, as bubbles internal circulation play a key role in controlling the transport of heat, mass, momentum through the interface \cite{ayyaswamy1990effect,Gissinger2021}, we characterized the mixing in the bubbles by studying the turbulent kinetic energy of the bubbles.
We observe a clear action of density and viscosity in modulating the turbulent kinetic energy of the bubbles.
In particular, a decrease of the bubble density or an increase of the bubble viscosity lead to a remarkable decrease of the turbulent kinetic energy levels in the bubbles.

\begin{acknowledgments}
We acknowledge ISCRA for awarding us access to Marconi-KNL (Project ID: HP10BOR3UN, 10M core hours and PRACE for awarding us access to HAWK at GCS@HLRS, Germany.
FM gratefully acknowledges funding from the MSCA-ITN-EID project \textit{COMETE} (project code 813948).
\end{acknowledgments}

\appendix
\section{Detection of coalescence and breakage events}
\label{appa}

In the simulations presented in the manuscript, topological changes are implicitly described by the phase-field method and thus no closure models are required to describe coalescence and breakage events.
To compute the coalescence and breakage rates, we use an algorithm that rely on the analysis of bubbles trajectories and bubbles volumes to identify topological modifications of the interface.

The input data needed are: the position of the center of mass of each droplet, identified by the subscript $i$, at the current time step, $\mathbf{x}^n_i$, the velocity of the center of mass of each droplet at the current time step, $\mathbf{u}^n_i$, and the position of the center of mass of each droplet at the following time step, $\mathbf{x}^{n+1}_i$.
These quantities are calculated for each droplet $i$ and are defined as:
\begin{equation}
\mathbf{x}^n_i=\frac{1}{V^n_i} \int_{V^n_i} \mathbf{x}^n_i \textnormal{d} V \, ;
\end{equation}
\begin{equation}
\mathbf{u}^n_i=\frac{1}{V^n_i} \int_{V^n_i} \mathbf{u}^n_i \textnormal{d} V \, ;
\end{equation}
\begin{equation}
\mathbf{x}^{n+1}_i=\frac{1}{V^{n+1}_i} \int_{V^{n+1}_i} \mathbf{x}^{n+1}_i \textnormal{d} V \, ,
\end{equation}
where the integral is computed over the volume $V_i$ of each droplet. The apices $n$ and $n+1$ identify respectively the current and the following time step; the elapsed time between the two time steps is $\Delta T$. 
In the first step the estimated position of each droplet at the following time step is computed as:
\begin{equation}
\mathbf{x}^{n+1}_{est,i}=\mathbf{x}^n_i+\Delta T \mathbf{u}^n_i\, .
 \end{equation}
To better explain the technique employed to detect translations, breakages and coalescences some examples have been reported in figure~\ref{fig: brt}.
For each droplet we compute the estimated position at the following time step $\mathbf{x}_{est,i}^{n+1}$, and we search for the closest bubble at the following time step; at this step some droplets at the following time step may be left out (they are not the closest droplet to any estimated droplet position). This step corresponds to figure~\ref{fig: brt}($a$): the estimated position of droplet $T_n$ is calculated (red semi-transparent bubble) and the closest bubble at the following time step is found out (droplet $T_{n+1}$).
In the following stage breakage and coalescence events have to be sorted out from these data.\\
A breakage is detected whenever a droplet in $\mathbf{x}^{n+1}$ has no parent droplet: according to figure~\ref{fig: brt}($b$) bubble $B_{n+1,2}$ has no parent bubble, thus it originated from a breakage event. 
Once a breakage event is identified, the algorithm searches for the the closest droplet to the bubble $B_{n+1,2}$ at time step $n+1$; in this case droplet $B_{n+1,1}$ is found. 
It is then assumed that droplet $B_n$ (whose estimated position is the closest to droplet $B_{n+1,1}$) breaks apart into droplets $B_{n+1,1}$ and $B_{n+1,2}$.
Once all breakages have been detected, the algorithm looks for coalescence events. 
A coalescence event is detected whenever two separate droplets at time step $n$ are assigned to the same droplet at time step $n+1$. 
In particular, referring to figure~\ref{fig: brt}($c$) bubbles $C^{n}_i$ and $C^{n}_j$ are both assigned to bubble $C^{n+1}_i$, as it is the closest one to their estimated position.
\begin{figure}
\begin{center}
\setlength{\unitlength}{0.0025\columnwidth}
\begin{picture}(230,250)
\put(0,5){\includegraphics[width=0.60\columnwidth, keepaspectratio]{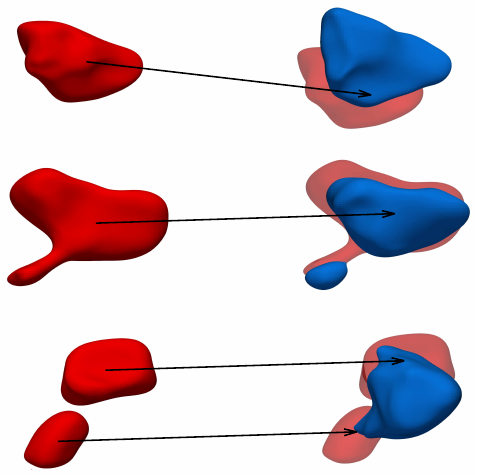}}
\put(-30,230){($a$)}
\put(-50,210){Translation}
\put(-30,155){($b$)}
\put(-47,135){Breakage}
\put(-30,68){($c$)}
\put(-53,48){Coalescence}
\put(42,232){$T_{n}$}
\put(202,232){$T_{n+1}$}
\put(44,155){$B_{n}$}
\put(195,106){$B_{n+1,1}$}
\put(150,90){$B_{n+1,2}$}
\put(12,60){$C_{n,1}$}
\put(-5,25){$C_{n,2}$}
\put(225,25){$C_{n+1}$}
\end{picture}
\caption{Possible cases considered for the algorithm: panel ($a$) corresponds to a translation, panel ($b$) to a breakage and panel ($c$) to a coalescence. 
Red bubbles are at the current time step ($n$), while blue bubbles are at the next time step available ($n+1$). 
Semi-transparent bubbles show the estimated position, $\mathbf{x}^{n+1}_{i,est}$. 
Arrows show the trajectory of the bubbles, $\Delta T \mathbf{u}^n_i$.}
\end{center}
\end{figure}
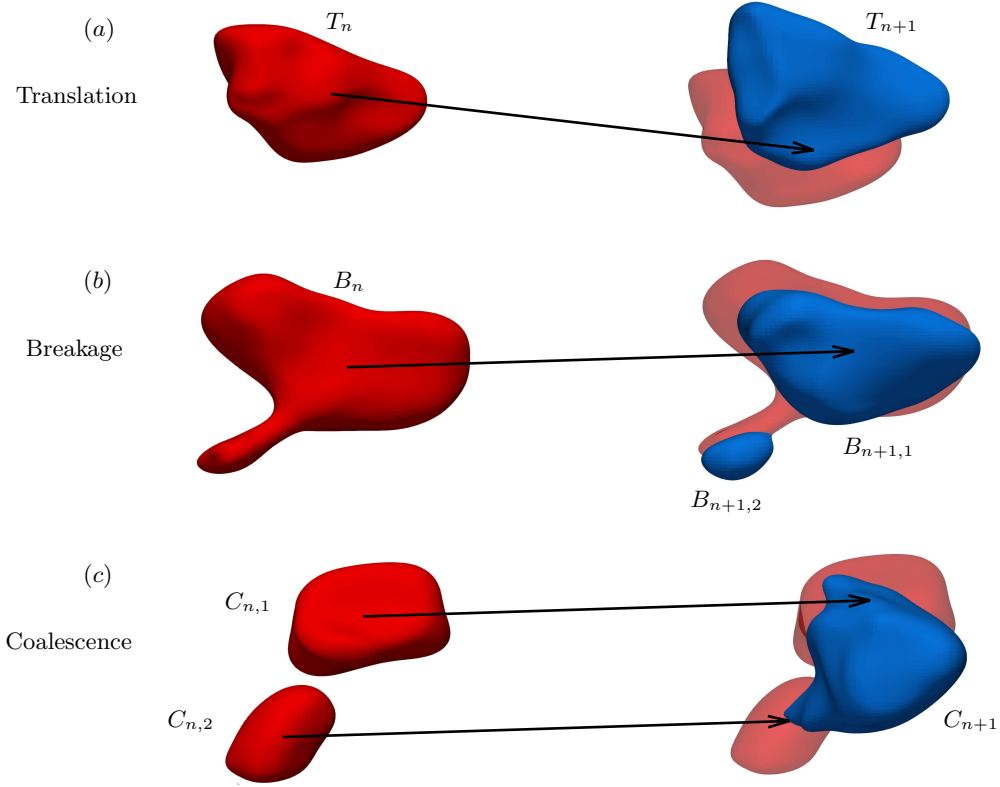
\begin{figure}
\begin{center}
\includegraphics[width=0.6\columnwidth]{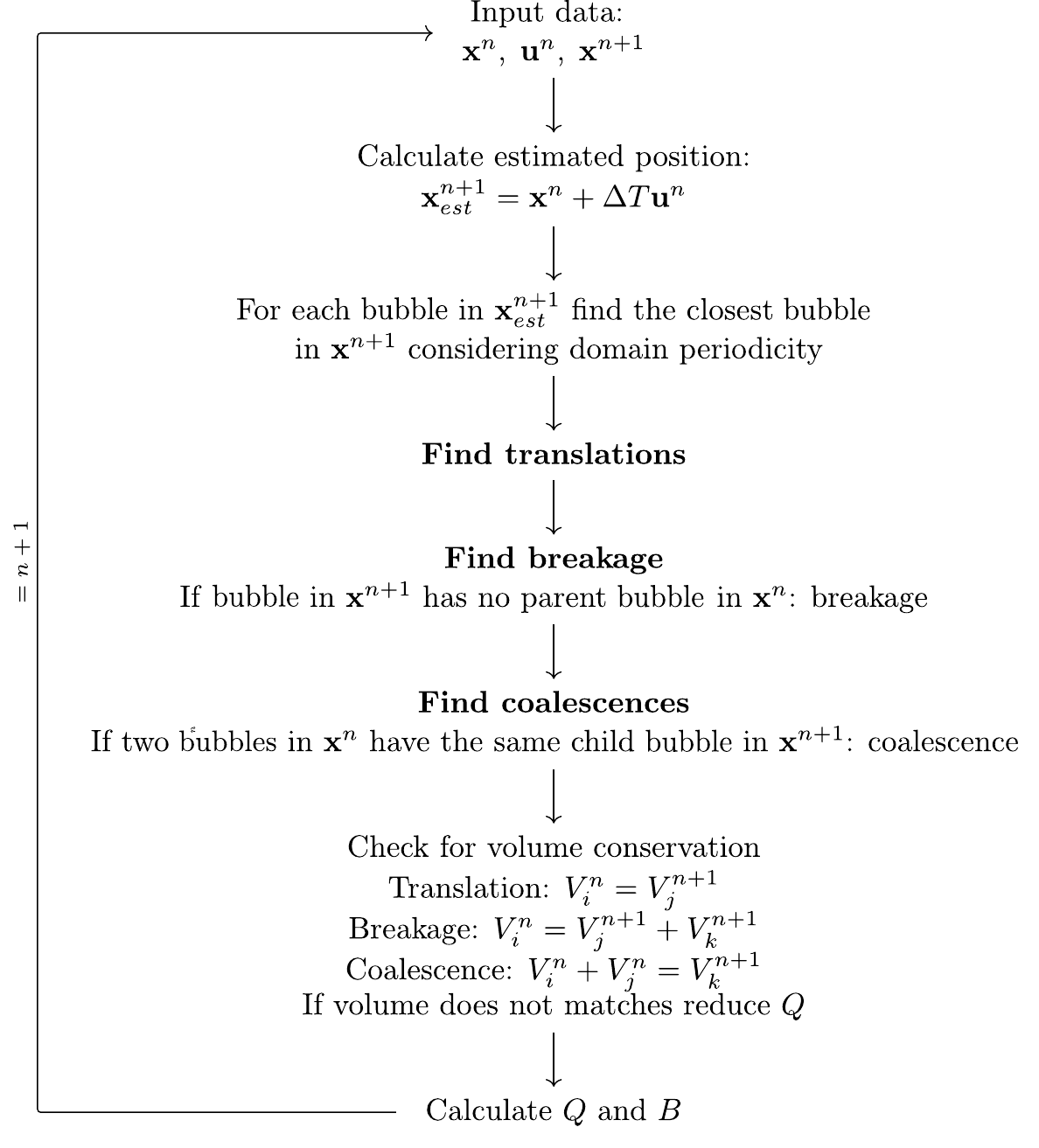}
\end{center}
\caption{Flow chart of the algorithm used to detect breakage and coalescence events in the post-processing of the simulations.}
\label{fig: brt}
\end{figure}
So far, only kinematic criteria have been used to determine the trajectory and eventual interactions (coalescences and breakages) of each bubble. 
Once all the trajectories at the present time step have been determined, the quality index and the balance are computed. 
In particular, the quality index, $Q$, is initialized at the beginning of the time step to the number of droplets at the current time step, $N_n$; every time volume is not conserved (within a certain small threshold) in all the translation, breakages and coalescences, the quality index is reduced by one. 
At the end of the time step, it is normalized by $N_n$. 
Recalling the examples of figure~\ref{fig: brt}, three checks on the volume conservation are performed depending on the type of event:
\begin{equation}
\begin{cases}
V_{T_n}= V_{T_{n+1}} \pm\varepsilon & \textnormal{for translations} \\
V_{B_n}= V_{B_{n+1,1}}+V_{B_{n+1,2}} \pm\varepsilon  & \textnormal{for breakages} \\
V_{C_{n,1}}+V_{C_{n,2}}= V_{C_{n+1}} \pm\varepsilon  & \textnormal{for coalescences} \\
\end{cases}  \; .
\end{equation}
To account for numerical errors that could occur in the calculation of the volume of each bubble (that would strongly reduce the quality index of the matching), a small tolerance $\varepsilon$ (of the order of few percents of the volume of parent droplet) is used when checking for volume conservation.\\
The second parameter controlling the quality of the calculated trajectories is the balance, $B$. 
The total number of bubbles at each time step is known: $N_n$ at the current time step and $N_{n+1}$ at the following one available. 
Once the number of breakage and coalescence events is known the balance can be calculated as:
\begin{equation}
B=N_{n+1}-(N_n+N_b-N_c)  \, ,
\end{equation}
where $B$ and $N_c$ are respectively the number of breakage and coalescence events that occur between time steps $n$ and $n+1$.
The number of droplets at the current time step, $N_n$, is increased whenever a droplet undergoes breakage into two bubbles and is decreased whenever two bubbles coalesce into one bubble. Here we make the assumption that all breakages are binary breakages and all coalescences involve only two parent droplets at a time.
Thus, considering these two parameters, a fair matching of the trajectories is obtained with a quality index $Q\to1$ and a balance $B=0$. 
This means that the volume is always matched (quality index never or rarely reduced) and no bubble is left out (balance equal to zero).\\
Finally, once known the number of coalescence and breakage rates that occur between each time step $n$ and $n+1$, the coalescence and breakage rates, $\dot{N_c}$ and $\dot{N_b}$, can be computed by counting the overall number of coalescence or breakage occurring in the temporal window $\Delta t^+$. 
Note that the temporal window used to track the trajectories of the bubbles is smaller than the temporal window used to compute the rates.
The present algorithm considers only binary breakages and coalescences events.
This assumption is not particularly limiting, as binary breakages/coalescences have the highest probability of occurrence \cite{Andersson2006,Maass2007,AziziT_2011}.
This assumption is also confirmed by the simulations performed: the quality index never reduces below 0.85 (so the volume is matched for at least 85\% of all the translation, breakage and coalescence events) and at most few droplets are left unmatched (the balance is almost zero).

\section{Influence of grid resolution on coalescence and breakage rates}
\label{appb}

\begin{figure}
\begin{center}
\includegraphics[width=0.6\columnwidth]{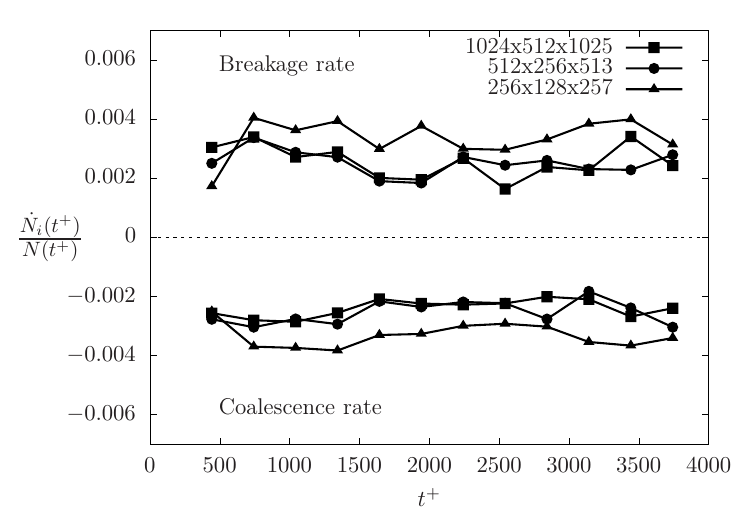}
\end{center}
\caption{Coalescence and breakage rates obtained using three different grid resolutions: $N_x \times N_y \times N_z =256 \times 128 \times 257$ (triangles), $N_x \times N_y \times N_z =512 \times 256 \times 513$ (circles) and $N_x \times N_y \times N_z =1024 \times 512 \times 1025$ (squares). 
The results refer to the case $We=3.0$, $\rho_r=1.000$ and $\eta_r=1.00$.}
\label{convergence_fig}
\end{figure}

To evaluate the influence of grid resolution on coalescence and breakage rates, we perform two additional simulations: one with a coarser grid resolution ($N_x \times N_y \times N_z =256 \times 128 \times 257$) and one with a more refined grid resolution ($N_x \times N_y \times N_z =1024 \times 512 \times 1025$).
The three simulations consider the same given case: $We=3.00$, $\rho_r=1.000$ and $\eta_r=1.00$.
As the grid resolution is changed, the Cahn number has been also adjusted accordingly (from $Ch=0.04$ for the coarser grid down to $Ch=0.01$ for the finer grid).
We compare the coalescence and breakage rates obtained from the three different grid resolutions in figure~\ref{convergence_fig}.
We can observe that for all the grid resolutions considered, the trend reported is similar and for all simulations, after an initial transient, both rates set to an equal value (in magnitude).
Analyzing the value of the rates obtained, we can notice that some differences are present between the coarser grid resolution (triangles) and the intermediate grid resolution (circles).
However, these differences become marginal when the intermediate grid resolution (circles) and the refined grid (squares) results are compared.
Overall, present results suggest that the mesh employed is sufficient to investigate breakage/coalescence dynamics.

\bibliography{totalbib.bib}

\end{document}